\documentclass{ametsocv6.1} 

\usepackage{hyperref}
\hypersetup{
   colorlinks = true,
   urlcolor   = blue,
   citecolor  = blue,
}
\usepackage{soul}
\setstcolor{red}
\usepackage{multirow}
\usepackage[position=top]{subfig}
\definecolor{RoyalBlue}{rgb}{0.2549,0.4118,0.8824}

\title{Turbulence in stratified rotating topographic wakes}

\authors{Jinyuan Liu$^{a}$, 
Pranav Puthan$^{a}$,
and Sutanu Sarkar$^{a,b}$ \correspondingauthor{Sutanu Sarkar, ssarkar@ucsd.edu}  
}

\affiliation{{$^a$}{Mechanical and Aerospace Engineering, University of California San Diego, La Jolla, CA 92093, USA}\\
{$^b$}{Scripps Institution of Oceanography, La Jolla, CA 92037, USA} 
}

\abstract{
Turbulence generation mechanisms in stratified, rotating flows past three-dimensional (3D) topography remain under-explored, particularly in submesoscale (SMS) regimes critical to geophysical applications. Using turbulence-resolving large-eddy simulations, we systematically dissect the interplay of stratification and rotation in governing the dynamics of wake turbulence. Our parametric study reveals that turbulent dissipation in the near wake is dominated by two distinct instabilities: (1) vertical shear-driven Kelvin-Helmholtz instability (KHI), amplified by oblique dislocation of K{\'a}rm{\'a}n vortices under strong stratification, and (2) centrifugal/inertial instability (CI), which peaks at intermediate rotation rates (Rossby number order unity, SMS regime) and relatively weaker stratification. Notably, strong rotation dampens vertical shear and weakens KHI-driven turbulence, while strong stratification imposes smaller vertical length scales that restrict CI-driven turbulence.  By quantifying dissipation across a broad parameter space of stratification and rotation, predictive relationships between the environmental parameters and instability dominance  {are} established. These findings highlight the regime dependence of instability mechanisms and may inform targeted observational campaigns and numerical models of oceanic and atmospheric wakes.
}  

\begin{document}

\maketitle

\section{Introduction} \label{intro} 


Observations of the upper ocean have revealed complex turbulent wakes and shed vortices, e.g., from headlands and island chains \citep{chang2013kuroshio,chang2019observations,mackinnon2019eddy,zeiden2021broadband,merrifield2019island,wynne2022measurements}.
Submerged topography in the deep ocean is also replete with three-dimensional features.  Individual seamounts in a chain and three-dimensional hills are associated with wake eddies, internal waves and turbulence. Even  a ridge that is two-dimensional at the large scale has three-dimensional features at the submesoscale (1 to 10 km in the horizontal). It has been hypothesized that steep seamounts and hills act as stirring rods  and constitute an important route to mixing since the energy of the incident current is a continuous reservoir of kinetic energy and the  stratified fluid carried by it is a continuous reservoir of potential energy. But, our knowledge of  turbulence and turbulent mixing by wakes in the deep ocean is severely limited by the scarcity of direct observations and the absence of non-hydrostatic large eddy simulations (LES) that can resolve these aspects of the flow. The overarching goal of high-fidelity numerical studies is to improve understanding and modeling of topographic wakes in the context of small-scale ocean turbulence and mixing, and the transport of water masses by large-scale coherent wake eddies. 

Recent studies \citep{johnston2019energy,mackinnon2019eddy,rudnick2019understanding} during the Flow Encountering
Abrupt Topography (FLEAT) initiative have shed light on the eddy field surrounding Palau Island which lies
in the path of the North Equatorial Current (NEC). Cyclonic and anti-cyclonic eddies spanning a
wide range of Rossby numbers ($Ro \approx 0.3$--$30$) are observed in the lee of the island with diameters
ranging from 1-10 km (submesoscale, SMS) to several hundred km (mesoscale, MS), which are comparable to the width
of the island. 

Direct measurements of microstructure in the Palau wake from a glider survey \citep{st2019turbulence} show that turbulent dissipation is enhanced in two different bands within the 200 m upper
ocean which includes strongly stratified thermocline waters. The elevated dissipation rate can be up to several orders of magnitude larger than the typical values observed in the stratified upper ocean layer \citep{st2019turbulence,mackinnon2019eddy,wijesekera2020observations,wynne2022measurements}, signifying the influence of the stirring role of the seamounts. Other notable examples of topographic wakes arise from interactions of the Kuroshio Current and the Gulf Stream with submerged topographies \citep{chang2013kuroshio,gula2016topographic}. 

It has been suggested in past theoretical work that the elevated dissipation can be driven by different types of instabilities \citep{thomas2013symmetric}, e.g., vertical shear instability (Kelvin-Helmholtz instability, KHI), SMS centrifugal/inertial instability (CI), and symmetric instability (SI).  
Examination of these instabilities has so far mainly considered the SMS eddies that are generated by baroclinically-unstable fronts. The role of these mechanisms in the dissipation of topographically-induced SMS eddies is poorly understood. 

Numerical studies of topographic wakes include the utilization of the hydrostatic regional oceanic modeling system (ROMS), such as idealized, isolated topography \citep{dong2007island,perfect2018vortex,srinivasan2021high} and realistic complex topography \citep{gula2016topographic,simmons2019dynamical}, and the {hydrostatic version} of the MIT-GCM such as \cite{liu2018numerical,nagai2021kuroshio,inoue2024numerical}. While hydrostatic simulations of topographic wakes capture well the general dynamics and provide parameterized turbulence and mixing, the details of instabilities and turbulence are not resolved. Recently, LES has been used to study the dynamics of topographic wakes, and has led to findings such as tidal synchronization of eddy shedding frequency \citep{puthan2021tidal}, elevated drag coefficient due to tidally modulated vortices \citep{puthan2022high} and coherent global modes in steady-current  wakes \citep{liu2024effect}. By applying such modeling to idealized three-dimensional topography whose non-dimensional parameters match the oceanic sites of interest, quantitative links between various metrics (turbulent dissipation, mixing, coherent vortex structures) and the governing non-dimensional parameters can be established, and further physical insights can be established,  enabling  applications to a broad range of three-dimensional topographic features in the ocean. 


Motivated by the need to understand and model topographic wake turbulence, we employ a suite of high-resolution LES with the objective being the characterization of  turbulence and mixing. The physical model is a three-dimensional (3D) conical topography sitting on the ocean bottom, as shown in figure 1 of \cite{liu2024effect}. The topography has a base diameter  {$D=500 \,{\rm m}$}, height $h=150 \, {\rm m}$ and a slope approximately $30^{\circ}$. This setting represents a model of a steep topography, such as seamounts. The fluid is linearly stratified and the coordinate system rotates at a constant rate ($f$-plane).  The flow impinging on the obstacle is a steady current and the focus of this work is on the comprehensive effects of stratification, rotation and 3D topography, {and how they affect the vortex dynamics and turbulent dissipation}. Such simplifications allow a focus on wake turbulence without the loss of generality. 

The rest of this paper is organized as follows: Section \ref{les} describes the LES numerical setting and the parameter selection, whereby the strengths of stratification and rotation are systematically varied. Section \ref{ins} examines the characteristics of the KHI and the CI and the resulting turbulent dissipation. In Section \ref{param}, the combined effects of stratification and rotation are considered, and turbulent dissipation is parameterized.  Section \ref{theend} provides a holistic summary and discussion of the results. 

\section{Numerical modeling} \label{les}

\subsection{Large-eddy simulations}

The flow is governed by the incompressible Navier--Stokes equations under Boussinesq approximation: 
\begin{align}
  \nabla \cdot \bm{u} & = 0 \\
  \frac{\partial \bm{u}}{\partial t} + \bm{u} \cdot \nabla  \bm{u} + \bm{f}_{c} \times 
   \bm{u}  & = -\frac{1}{\rho_0} \nabla p^* + \nabla \cdot \bm{\tau} + b {\bm{e}}_z \label{eqn:momentum}  \\
  \frac{\partial \rho}{\partial t} + \bm{u} \cdot \nabla  \rho  &= \nabla \cdot {\bm{J}_{\rho}} 
\end{align} where the viscous stress and the scalar flux are \begin{equation}
  \bm{\tau} = (\nu+\nu_{\rm sgs}) (\nabla \bm{u} + (\nabla \bm{u})^T
   ), \, {\bm{J}_{\rho}}  = (\kappa + \kappa_{\rm sgs}) \nabla \rho,  
\end{equation} 
the buoyancy  is $ b = - {\rho^* g}/{\rho_0}$.
Here $p^*$ and $\rho^*$ are the deviations from the hydrostatic and geostrophic balances, and $\nu$ and $\kappa$ are the molecular viscosity and diffusivity. The subscript `sgs' denotes the sub-grid scale contributions to the momentum and scalar transport,  and are modeled with the WALE (wall-adapted local eddy-viscosity; \cite{nicoud1999subgrid}) method in an LES approach. The Coriolis force involves the planetary rotation vector,  $\bm{f}_{ c} = f_{c} {\bm{e}}_z$,  {where the sign of the Coriolis frequency $f_c$ is negative (Southern Hemisphere), and the magnitude of $f_c$ is a constant as per the $f$-plane approximation, in the present work.}  {We opt to use a Southern Hemisphere rotation but the results are equally applicable to both hemispheres with the distinction between cyclones and anticyclones carefully made. }

The non-dimensional Froude, Rossby, and Reynolds numbers, \begin{equation} 
  Fr = \frac{U_{\infty}}{Nh} ,\, Ro = \frac{U_{\infty}}{|f_{c}| D}, \, Re = \frac{U_{\infty} D}{\nu}, 
\end{equation} are the key controlling parameters that will be systematically varied in this study,  {where the definition of $Ro$ involves only the magnitude of $f_c$.} 

The governing equations are solved with a finite-difference solver that has an immersed boundary formulation \citep{balaras2004modeling,yang2006embedded} to deal with  topography. Second-order central differences on a staggered grid are used to discretize the spatial derivatives, and a third-order Runge--Kutta scheme is used for time advancement. A fractional step method is used to obtain time-accurate divergence-free velocity fields and the resulting three-dimensional pressure Poisson equation is solved with a direct method. This solver is generally validated in the simulation of unstratified and stratified turbulent wakes \citep{pal2017direct,chongsiripinyo2020decay}. The setting is very similar to that of \cite{liu2024effect}, but the parameter space explored in the present work is substantially larger, and the focus is on turbulence and its generation mechanisms.  More details of the numerical setup can be found in Appendix \hyperref[sims]{A}.

\begin{figure*}[thb]
  \centering 
  \captionsetup[subfloat]{farskip=0pt,captionskip=1pt}
  \subfloat{{\includegraphics[width=0.75\linewidth]{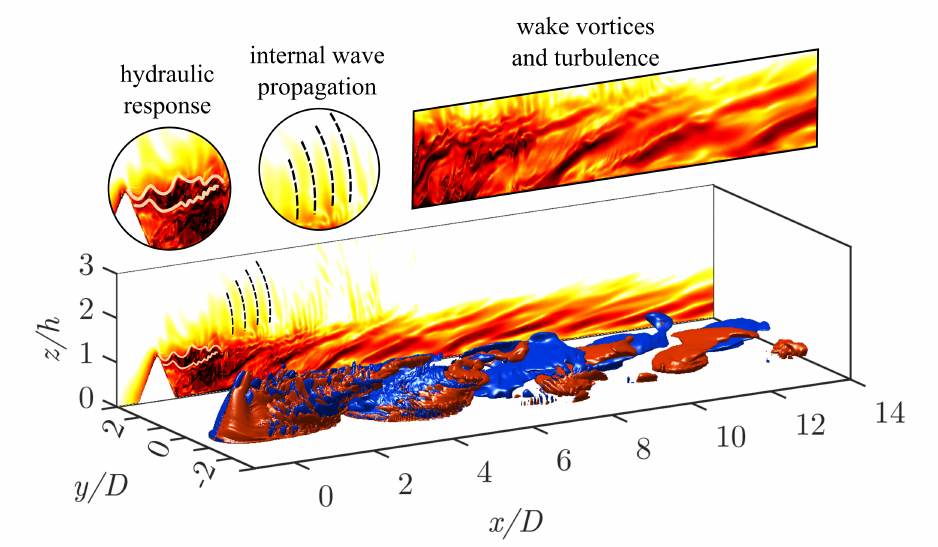}}}%
  \caption{ {Topographic wake generated by a conical seamount.  Red and blue contours show the isosurfaces of vertical vorticity, $\omega_z D/U_{\infty}=\pm1$, representative of oblique  vortex shedding structures. The sidewall projection shows the turbulent dissipation rate in logarithmic scale (for the colorbar see Fig.~\ref{dissp_rig}), at the same time instance and on the central plane ($y=0$). The upper panels show specific regions of the flow, featuring distinct physical mechanisms. The flow goes from left to right. Case Fr030Ro3.}} \label{flow}
\end{figure*}

\subsection{ {Flow features and parameter selection}} 

 {Topographic wakes are enriched by multi-physics and multi-scale interactions.  Fig.~\ref{flow} depicts the flow patterns. As the flow passes the obstacle, a hydraulic response is established near the top of the seamount resulting in downslope flow and an undulatory jet, both with strong vertical shear. Moreover, internal waves are transmitted into the stratified medium. In the wake, eddies of different scales ranging from large-scale coherent wake eddies to dissipative eddies coexist and interact.} 
 {At the large scale, the wake exhibits K{\'a}rm{\'a}n shedding patterns at different heights. It is intuitive to think that, in strongly stratified flows,  K{\'a}rm{\'a}n shedding is vertically decoupled into independent, stacked layers. However, the vortex shedding structures were found to share a vertically independent global frequency in the $Fr=0.15$ seamount wakes investigated by \cite{liu2024effect}. When rotation is not strong, the coherent structures are oblique `surfboards'. They lean into the streamwise direction and form a shallow angle to the horizon, as shown in Fig.~\ref{flow}. On one hand, the existence of these coherent structures is robust to the presence of near wake turbulence. On the other hand, the vortex and shear structures associated with them provide a breeding bed for various instabilities. }   

 {In this work, we are mainly concerned with wake turbulence, where several physical mechanisms can potentially lead to the destabilization of the wake and the breakdown into turbulence. Vertical shear that is sufficiently strong compared to $N$ will be unstable to  KHI, where the source of shear can be in between opposite-signed `surfboard' vortices or from the intensified shear of the hydraulic jet (HJ). In the horizontal directions, anticyclonic vortices and shear layers are both subject to CI.} These two key mechanisms -- KHI and CI -- are determined to be operative in  wake turbulence later on and  their parametric dependence is investigated.

\begin{table*}[htb]
  \begin{center}
  \begin{tabular}{l l l l c l l l c}
  \topline
   case & $Fr$ & $Ro$ & $Bu$ & $Re$ & $N$ (s$^{-1}$) &  $|f_{c}|$ (s$^{-1}$) & $U_{\infty}$ (m s$^{-1}$) & color \\
  \midline
  Fr007Ro015  & \multirow{5}{*}{0.075} & 0.15 & 4    & \multirow{5}{*}{40 000}  & $9.33 \times 10^{-4}$ & $1.40 \times 10^{-4}$ & $1.05 \times 10^{-2}$ &  \multirow{5}{*}{blue} \\
  Fr007Ro050  &                       & 0.5  & 44   &   & $9.33 \times 10^{-4}$ & $4.20 \times 10^{-5}$ & $1.05 \times 10^{-2}$ \\
  Fr007Ro075  &                       & 0.75 & 100  &   & $9.33 \times 10^{-4}$ & $2.80 \times 10^{-5}$ & $1.05 \times 10^{-2}$\\ 
  Fr007Ro1p5  &                       & 1.5  & 400  &   & $1.87 \times 10^{-3}$ & $2.80 \times 10^{-5}$ & $2.10 \times 10^{-2}$ \\
  Fr007Ro7p5  &                       & 7.5  & $1\times 10^4$  &  & N/A  & N/A  & N/A  \\
  \midline
  Fr015Ro015  & \multirow{5}{*}{0.15} & 0.15 & 1    & \multirow{5}{*}{20 000}  & $4.67 \times 10^{-4}$ & $1.40 \times 10^{-4}$ & $1.05 \times 10^{-2}$ & \multirow{5}{*}{green}  \\
  Fr015Ro075  &                       & 0.75 & 25   &   & $4.67 \times 10^{-4}$ & $2.80 \times 10^{-5}$ & $1.05 \times 10^{-2}$ \\
  Fr015Ro1p5  &                       & 1.5  & 100  &   & $9.33 \times 10^{-4}$ & $2.80 \times 10^{-5}$ & $2.10 \times 10^{-2}$ \\
  Fr015Ro3    &                       & 3    & 400  &   & $1.87 \times 10^{-3}$ & $2.80 \times 10^{-5}$ & $4.20 \times 10^{-2}$ \\
  Fr015Ro7p5    &                      & 7.5    & $2.5\times 10^3$ &   & N/A & N/A & N/A \\
  \midline
  Fr030Ro030  &  \multirow{5}{*}{0.3} & 0.30 & 1    & \multirow{5}{*}{10 000}& $4.67 \times 10^{-4}$ &   $1.40 \times 10^{-4}$ & $ 2.10 \times 10^{-2}$ & \multirow{5}{*}{red} \\
  Fr030Ro075  &                       & 0.75 & 6.25 &                        & $1.17 \times 10^{-3}$  & $1.40 \times 10^{-4}$ & $5.25 \times 10^{-2}$ \\
  Fr030Ro1p5  &                       & 1.5  & 25   &                        & $1.17 \times 10^{-3}$  & $7.00 \times 10^{-5}$ & $5.25 \times 10^{-2}$ \\
  Fr030Ro3    &                       & 3    & 100  &                        & $1.17 \times 10^{-3}$ & $3.50 \times 10^{-5}$ & $5.25 \times 10^{-2}$ \\
  Fr030Ro7p5    &                       & 7.5    & 625  &                      & N/A & N/A & N/A \\
  \midline
  Fr040Ro015  & \multirow{5}{*}{0.4}  & 0.15 & 0.14    & \multirow{5}{*}{7 500} &  N/A  & N/A & N/A & \multirow{5}{*}{yellow} \\ 
  Fr040Ro040  &                       & 0.40 & 1  &                        & $4.67 \times 10^{-4}$   & $1.40 \times 10^{-4}$ & $2.80 \times 10^{-2}$  \\
  Fr040Ro075  &                       & 0.75 & 3.5  &                        & $8.75 \times 10^{-4}$   & $1.40 \times 10^{-4}$ & $5.25 \times 10^{-2}$  \\
  Fr040Ro1p5  &                       & 1.5  & 14  &                         & $1.75 \times 10^{-3}$   & $1.40 \times 10^{-4}$ & $1.05 \times 10^{-1}$\\
  Fr040Ro4    &                       & 4    & 100 &                         & $1.46 \times 10^{-3}$   & $4.38 \times 10^{-5}$ & $8.75 \times 10^{-2}$ \\
  \botline
  \end{tabular}
  \caption{Parameters of the computational study which has 4 series corresponding to $Fr = 0.075$, 0.15, 0.30 and 0.40 with  $Ro$ varying within a series.
  Each series has  a fixed combination of $Fr$ and $Re$ such that the stratification scale based Reynolds number, $Re_N = U_{\infty}^2/\nu N = Fr Re (h/D)$, is a constant at $Re_N = 900$. 
   {Dimensional reference values for $N, |f_c|$ and $U_{\infty}$ are given for 
  }  buoyancy frequency in the range $0.5\times10^{-3}\, {\rm s^{-1}} \le N \le 0.5\times10^{-2}\, {\rm s^{-1}}$ and the Coriolis frequency in the range $2.5 \times 10^{-5} \le |f_c| \le  1.4 \times 10^{-4}$ (latitudes between 10$^\circ$ and 75$^\circ$),  {based on $D=500$ m and $h=150$ m. Cases with reference values not provided (marked as N/A) are still dynamically relevant with different values of $D$ and $h$ but do not have reasonable dimensional values for $D=500$ m and $h=150$ m. }
  } \label{table1}
  \end{center}
\end{table*}

\begin{figure}[thb]
  \centering 
  \captionsetup[subfloat]{farskip=0pt,captionskip=1pt} 
  \subfloat{{\includegraphics[width=0.55\linewidth]{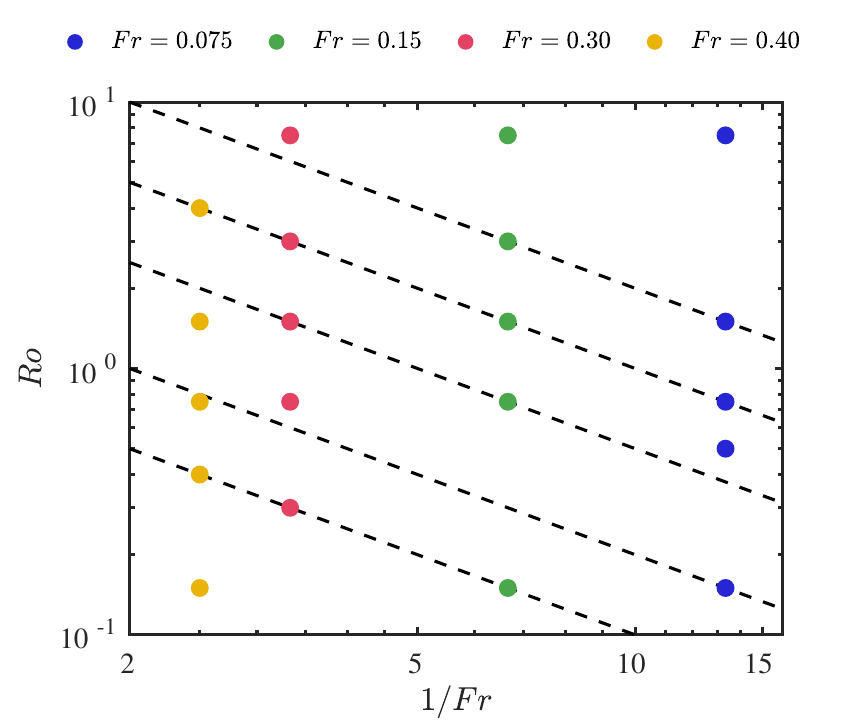}}}%
  \caption{Parameter space spanned by $(Fr,Ro)$. The horizontal axis is the inverse Froude number, $Fr^{-1}$, and the vertical axis is the Rossby number, $Ro$. Dashed lines represent constant Burger numbers, $Bu=(Ro/Fr)^2 = 1, 4, 25, 100, 400$, increasing from the lower left to the upper right. For each $Fr$, the four cases enclosed by $1\le Bu \le 400$, will be analyzed in detail. } \label{para_space} 
\end{figure} 

In order to cover a wide range of combinations of stratification and rotation, a parameter sweep is conducted in the $Fr$--$Ro$ space, as shown in Table \ref{table1} and Fig.~\ref{para_space}. The non-dimensional parameters are selected as follows.  

Although the stratification in the deep ocean has seasonal and geographical variability, a buoyancy frequency of $N=0.5\times10^{-3}\, {\rm s^{-1}}$ can be taken as representative. A mountain height of 1 km and a current speed between $0.2\,{\rm m\,s^{-1}}$ and $0.02\,{\rm m\,s^{-1}}$ leads to $Fr$ between 0.4 and 0.04. Accordingly, four values of $Fr$, from 0.075 to 0.40, that span half an order of magnitude, are selected. All cases lie in the flow-around regime \citep{hunt1980experiments,chomaz1993structure}, and they represent moderate to relatively strong stratification. The focus is on steep underwater topography that are known sites of elevated mixing and, accordingly,  {a representative value of} $h/D = 0.3$  {corresponding to a slope of approximately 30$^{\circ}$ is chosen}.  {Here, `steep' refers to the following dynamical aspects:  (1) the terrain slope exceeds typical angles of oceanic  internal wave propagation,  and (2) a substantial part of the wake is in the flow-around, vortex-shedding regime.}

Regarding rotation, the SMS regime ($Ro = O(1)$ where rotation influences the flow but is not sufficiently strong to dominate the dynamics) has attracted much recent interest.
 For each $Fr$, we select an overall range of $Ro$ from 0.15 to 7.5, which spans more than an order of magnitude, and is centered on the SMS while also including the limits of small MS and weakly rotating regimes.


The $Fr$--$Ro$ parameter space also extends the scope of previous work employing ROMS. The work of \cite{perfect2020energetics1} focused on large topographies in the MS with $Fr, Ro \sim O(0.01-0.1)$. Similarly, \cite{srinivasan2021high} studied strong rotation and strong stratification ($Fr = 0.02, \, 0.025 \le Ro \le 1$, in the present definition). In the limits of strongly rotating and/or stratified flows, the hydrostatic assumption in ROMS generally works well {for the large-scale motions} and both stratification and rotation act as stabilizing factors for smaller-scale motions \citep{perfect2020energetics1}. {When hydrodynamic instabilities and turbulence are of interest as is the case here, non-hydrostatic simulations are required to resolve them and, thus, elucidate their qualitative and quantitative  dependence on $(Fr,Ro)$. }


There is an additional (but not independent) non-dimensional number, the Burger number ($Bu=(Ro/Fr)^2 = (N h/f_c D)^2$), commonly used to describe the relative importance of rotation to stratification in geophysical flows. When $Bu\le O(1)$, rotation dominates 
and when $Bu>O(10)$, the effect of rotation is relatively small compared to the dominance of stratification. 
The range of $1 \le Bu \le 400$, shown by the four diagonal dashed lines (each has a constant value of $Bu$)  in Fig.~\ref{para_space}, will be examined in detail. 

\begin{figure*}[thb]
  \centering 
  \captionsetup[subfloat]{farskip=-2pt,captionskip=0pt}
  \subfloat{{\includegraphics[width=.6\linewidth]{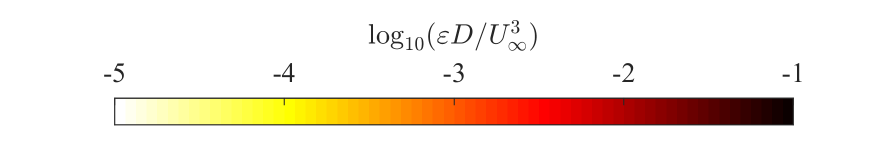}}} \hfill \\
  \subfloat{{\includegraphics[width=0.5\linewidth]{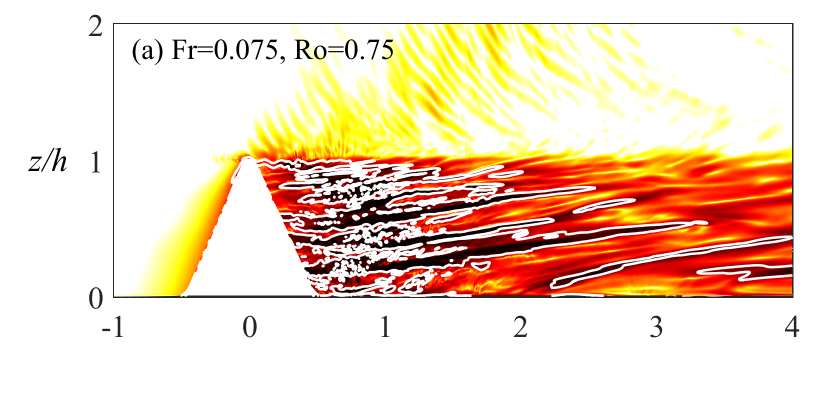}}}%
  \subfloat{{\includegraphics[width=0.5\linewidth]{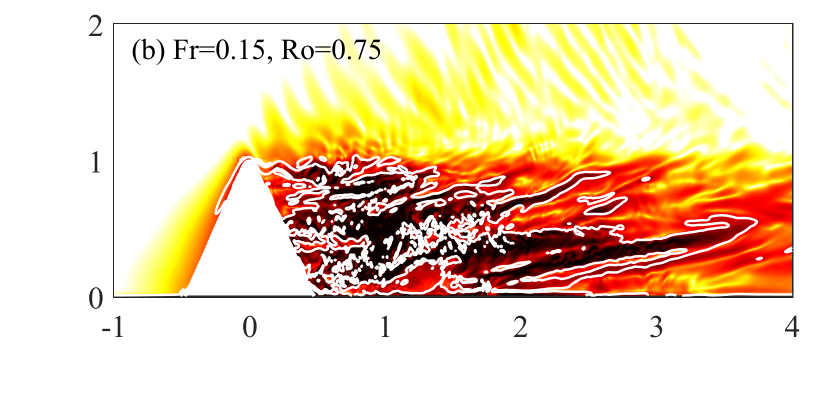}}}%
  \\ 
  \subfloat{{\includegraphics[width=0.5\linewidth]{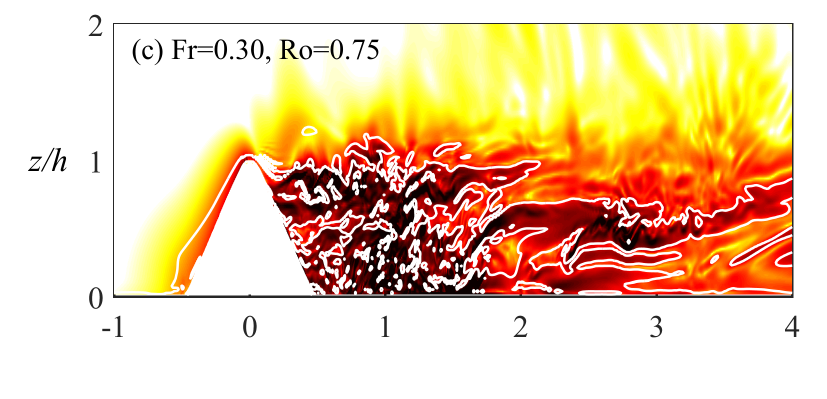}}}%
  \subfloat{{\includegraphics[width=0.5\linewidth]{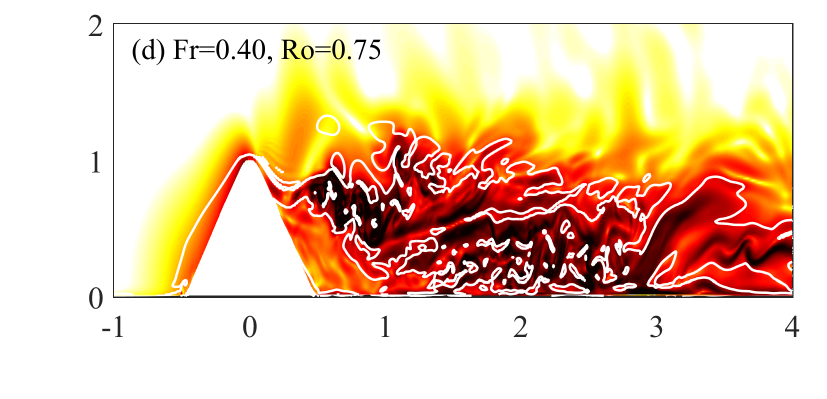}}}%
  \\
  \subfloat{{\includegraphics[width=0.5\linewidth]{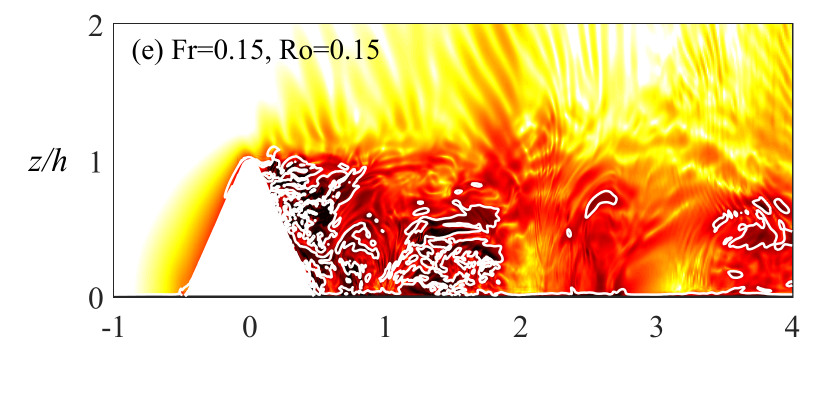}}}%
  \subfloat{{\includegraphics[width=0.5\linewidth]{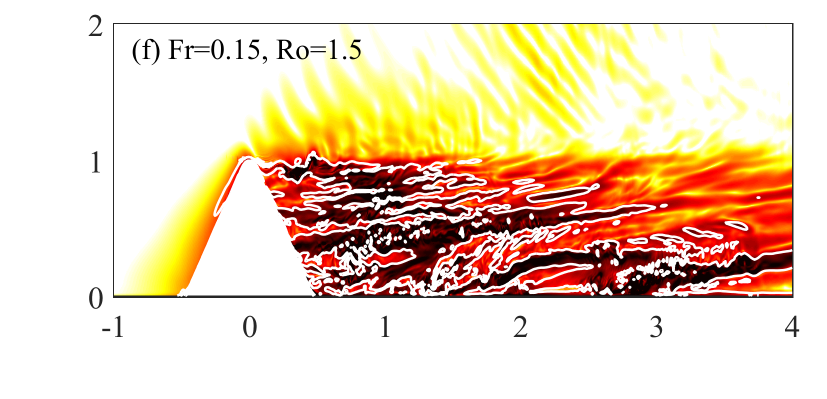}}}%
  \\
  \subfloat{{\includegraphics[width=0.5\linewidth]{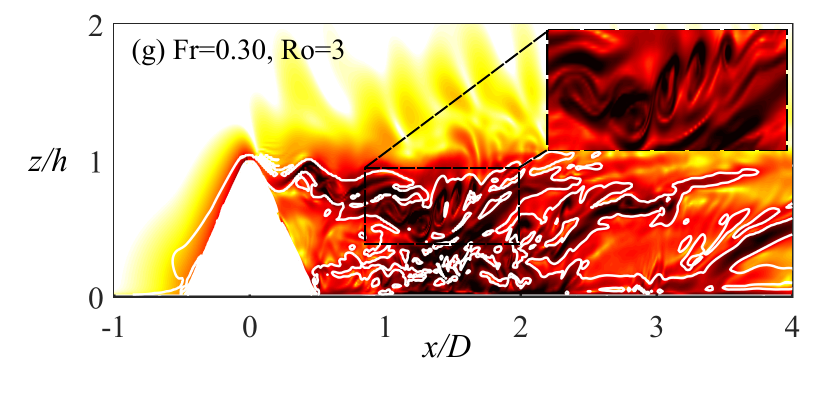}}}%
  \subfloat{{\includegraphics[width=0.5\linewidth]{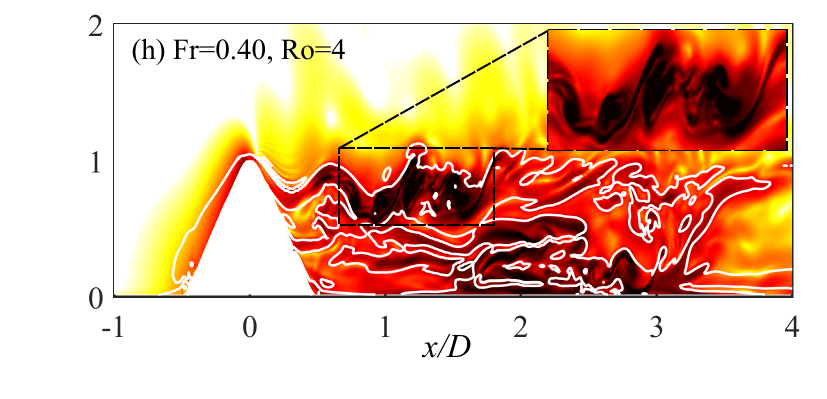}}}%
  \caption{Instantaneous dissipation rate $\varepsilon$ in the center plane ($y=0$), with white isolines of instantaneous values of $Ri_g=1/4$ overlaid on the top. (a-d) $Fr=0.075, 0.15, 0.30, 0.40$ and $Ro=0.75$. (e,f) $Fr=0.15$,  and $Ro=0.15, 1.5$, respectively. (g) $Fr=0.30, Ro=3$. (h) $Fr=0.40,Ro=4$. The insets of (g,h) are enlarged views of the KH rollups due to the hydraulic jet.  
  } \label{dissp_rig}
\end{figure*}

The chosen Reynolds number should be sufficiently high so that the flow becomes turbulent and as many instabilities are triggered as possible. At the same time, $Re$ cannot be too high so as to avoid  numerical dispersion errors arising from flow instabilities that are poorly resolved on the chosen grid.
At $Fr=O(0.1)$, the regime studied here, the wake is in the flow-around vortex shedding regime, where the horizontal components dominate the turbulent kinetic energy (TKE). However, there are vertical structures and (oblique) dislocations (see Fig.~\ref{dissp_rig}) of these horizontal motions, leading to strong vertical shear and the potential for KHI. This scenario is a generic feature of vertically stratified horizontal shear flows \citep{billant2000experimental,basak2006dynamics} and such a turbulence generation mechanism is referred to as a shortcut in the transition \citep{deloncle2008nonlinear,waite2008instability}, since the flow is originally sheared in the horizontal directions.   


The normal-mode KHI requires at least one point in the flow to have a gradient Richardson number less than 1/4 \citep{miles1961stability,howard1961note}.  {From Fig.~\ref{flow}, it can be seen that a surfboard structure lies on top of another with oppositely-signed vorticity (near $x/D=4$). This scenario constitutes an oblique dislocation. When the direction of spanwise wake flapping, indicated by the sign of K{\'a}rm{\'a}n vortices, is opposite across a vertical distance, intense vertical shear is generated.} Using the viscous length scale between dislocated layers measured in \cite{basak2006dynamics}, approximately $l_d = 15 \sqrt{\nu/N}$, and a velocity difference $2U_{\infty}$ between oppositely flapping  {wake vortices separated by this distance}, the resulting vertical shear is $S_v=2U_{\infty}/l_d$.  {The Reynolds number at which this shear is marginally stable, such that $Ri_g=N^2/S_v^2=1/4$, is $Re_N=U_{\infty}^2/\nu N = 225$. }



Thus, $Re_N=U_{\infty}^2/\nu N=FrRe(h/D)$, instead of $Re$ or $Fr$ alone, serves as the {\it a priori} indicator of transition to turbulence through KHI in the present wakes. It can also be interpreted as the Reynolds number of 
 {turbulent eddies whose size does not exceed the stratification length scale $\sim U_{\infty}/N$}.  {Such an estimate has the following implications. First, the Reynolds number required to have turbulence increases with stratification and can be quite large at strong stratification.} For example, for $Fr = 0.15$, the constraint of $Re_N >225$ requires a molecular Reynolds number of $Re>5000$, which is higher than what many laboratory experiments and  numerical simulations of the large-eddy and direct numerical simulation (LES and DNS) classes have reached. 
 {Second,  if a similar dynamic range of  turbulence is desired for a series of experiments, one also needs to consider matching a lower $Fr$ with a higher $Re$ and vice versa so that $Re_N$ is comparable, in addition to fixing one and varying another. }
We base the selection of the Reynolds number on such an estimate and   {hold $Re_N=900$ constant across all cases as an {\it a priori} choice of a buoyancy-based  Reynolds number. The value of  $Re_N=900$ is four times the estimated critical value so as to ensure fully-developed  turbulence.} It will be shown {\it a posteriori} that  KHI is active in all cases. The highest Reynolds number is up to $Re=40\,000$ at $Fr=0.075$.  
A Reynolds number sensitivity study is provided in Appendix \hyperref[re_sens]{B} to show that  turbulent dissipation statistics approach $Re$-independence in the present simulations.

Each case in Table \ref{table1} is run for more than two flow-throughs to eliminate transient effects before data collection, which spans approximately five flow-throughs ($100D/U_{\infty} $ or 25 vortex shedding cycles). 





\section{Dissipation features  and their link to flow instabilities
} \label{ins}

\subsection{Turbulence statistics and quantification} 
Definitions and notations that are used in the statistical analysis are as follows. An instantaneous signal is decomposed into  {the mean} and fluctuation as \begin{equation}
  \varphi = \langle {\varphi} \rangle + \varphi', 
\end{equation}  {where the bracket $\langle \cdot \rangle$ denotes time average.}   {The ensemble size for the time average is $N_t \approx 375$ when snapshots of 2D planes are analyzed, and is $N_t \approx 30$ when snapshots of 3D boxes are analyzed}.  The turbulent kinetic and potential energy (TKE and TPE) are  \begin{align}
  k & = \frac{1}{2} ( \langle {u'^2} \rangle + \langle {v'^2} \rangle  + \langle {w'^2} \rangle  ); \;
  k_{\rho}   = \frac{1}{2N^2}  \langle {b'^2} \rangle, 
\end{align} and the instantaneous dissipation rates are  \begin{align}
  {\varepsilon} & =  (\nu+\nu_{\rm sgs}) \frac{\partial u'_i}{\partial x_j} \frac{\partial u'_i}{\partial x_j}; \;    {\varepsilon}_{\rho}  = \frac{1}{N^2}  (\kappa+\kappa_{\rm sgs}) \frac{\partial b'}{\partial x_j} \frac{\partial b'}{\partial x_j},\label{e1} 
\end{align} whose time averages are the TKE and TPE dissipation, $\langle \varepsilon \rangle $ and $\langle {\varepsilon}_{\rho} \rangle$, respectively.  {Similarly, the instantaneous dissipation rates of the total kinetic and potential energy are denoted with tildes to distinguish from \eqref{e1}:}  \begin{align}
  \tilde{\varepsilon} & =  (\nu+\nu_{\rm sgs}) \frac{\partial u_i}{\partial x_j} \frac{\partial u_i}{\partial x_j}  ;\; 
  \tilde{\varepsilon}_{\rho}  = \frac{1}{N^2}  (\kappa+\kappa_{\rm sgs}) \frac{\partial b}{\partial x_j} \frac{\partial b}{\partial x_j}.  \label{eps_def}
\end{align} 


  




\begin{figure*}[t!]
  \centering 
  \captionsetup[subfloat]{farskip=0pt,captionskip=1pt}
  \subfloat{{\includegraphics[width=0.25\linewidth]{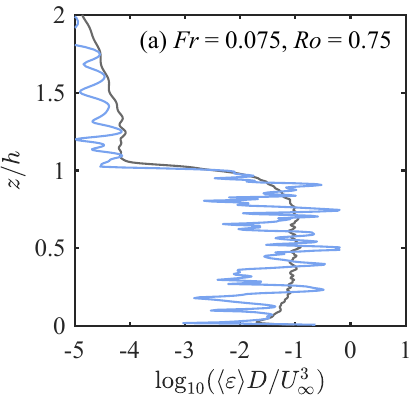}}}%
  \subfloat{{\includegraphics[width=0.25\linewidth]{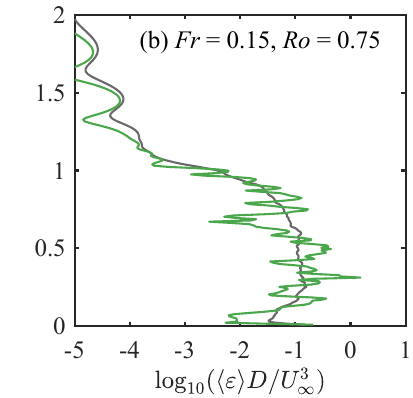}}}%
  \subfloat{{\includegraphics[width=0.25\linewidth]{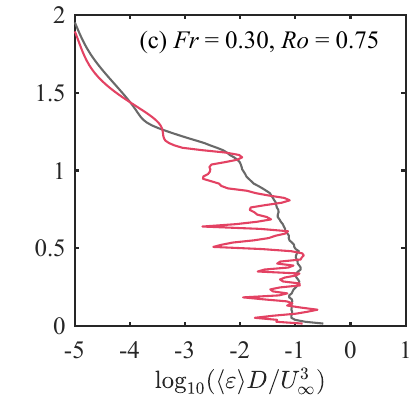}}}%
  \subfloat{{\includegraphics[width=0.25\linewidth]{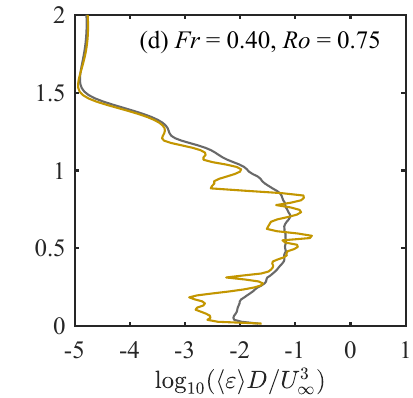}}}%
  \caption{Instantaneous dissipation rate $\varepsilon$ (colored) and its time-averaged $\langle {\varepsilon} \rangle$ (TKE dissipation, gray), probed on the line $y=0, x/D=1$. Time instances are  the same as those in Fig.~\ref{dissp_rig}(a-d). 
  } \label{dissp_x1_y}
\end{figure*}

\subsection{Contribution of the vertical shear instability (KHI)}



Figure \ref{dissp_rig} shows the normalized instantaneous TKE dissipation rate, $\varepsilon D/U_{\infty}^3$, enclosed by the white contours of the local gradient Richardson number indicator, $Ri_g=1/4$, defined as \begin{equation}
  Ri_g = \frac{(\partial_z b)^2}{(\partial_z u)^2 + (\partial_z v)^2}.  \label{rig_def}
\end{equation} Here $u,v$ and $b$ are instantaneous quantities. A value of $Ri_g=1/4$ typically indicates a marginal instability state due to vertical shear. It is clear that the shear-unstable regions coincide with the strongest dissipation, suggesting that KHI is active. 



Comparing Figs.~\ref{dissp_rig}(a-d), which are at the same $Ro=0.75$, it can be seen that strong localized dissipation has a similar magnitude at different $Fr$. Furthermore, as stratification increases ($Fr$ decreases), the number of oblique layers increases and the thickness of the layers also reduces. These spatial structures of dissipation align well with the vortex structures in these wakes when rotation is not dominantly strong \citep{liu2024effect}, which are indeed slanted 3D coherent structures  instead of  dislocated stacks of pancake vortices as previously suggested. Without the dominance of rotation, {each individual vortex is better described by the tilted vortex model in \cite{boulanger2007structure,canals2009tilted} instead of pancake vortices}. Although tilted and pancake-shaped vortices are qualitatively different and each has distinct vorticity--density structures \citep{beckers2001dynamics,basak2006dynamics}, they are both associated with intensified vertical shear due to the flow layering -- a direct consequence of stratification.

Comparing Figs.~\ref{dissp_rig}(b,e,f), at $Fr=0.15$ and various $Ro$, it can be seen that the SMS cases $Ro=0.75, 1.5$ are similar while the MS case $Ro=0.15$ shows fewer large-$\varepsilon$ patches. At $Ro=0.75, 1.5$, layers of tilted coherent vortex structures are shed and the shear instability serves as the main contributor to turbulence. When rotation is strong ($Ro=O(0.1)$), vertical gradients are significantly reduced and 
columnar vortices emerge which, further downstream, advect as  stratified Taylor columns \citep{liu2024effect}. Figure \ref{dissp_rig}(e) shows turbulent dissipation associated with these columnar vortices, which is significantly weaker than the dissipation associated with the slanted layers in (b,f). The contrast indicates that the form of the coherent structures, in turn, influenced by rotation, influences turbulence intensity. This might be regarded as the indirect effect of rotation on turbulence. 




Another contributor to turbulence is the HJ that plunges below the mountain crest and also sets up a near-field internal wave response. {The topmost portion of the mountain, from its crest to $U_f/N$ below \citep{winters2012hydraulic}, participates in the jet and the lee wave, whose dominant vertical wavelength is $2 \pi U_m/N$ \citep{klymak2010high}. Here $U_f$ and $U_m$ are the freestream and mean velocity, respectively.}  The undulating jet and the adjacent wave region constitute a dissipation hotspot that becomes more pronounced with increasing $Ro$ and
$Fr$ of the examined cases. The flow in the hotspots breaks down through KHI and forms rolled-up billows that are evident in the insets of Figs.~\ref{dissp_rig}(g-h), corresponding to Fr030Ro3 and Fr040Ro4.
 


Moreover, in all cases, there is turbulent dissipation due to the unsteady internal waves propagating into the background, as shown by the yellowish colors above the top of the seamount, {albeit smaller than the dissipation in the wake} (note the logarithmic color scale in Fig.~\ref{dissp_rig}). While the wave dissipation represents a different process that is also important, wake dissipation will be the focus of this work. In the vertical center plane, TPE dissipation (not shown) has a spatial distribution that aligns well with TKE dissipation. {Accompanying movies of $\varepsilon$ for vertical center planes as shown in panels (a-d) can be found in the supplementary materials.}




\begin{figure*}[thb]
  \centering 
  \captionsetup[subfloat]{farskip=-6pt,captionskip=1pt}
  \subfloat{{\includegraphics[width=0.4\linewidth]{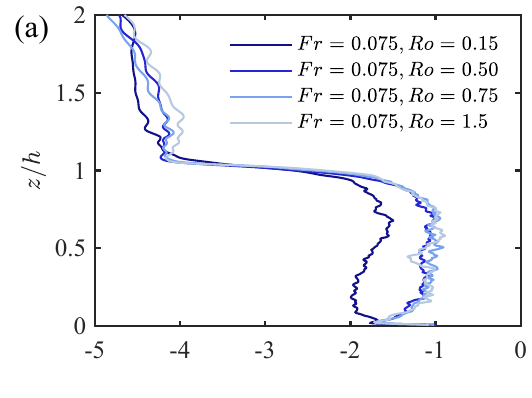}}}%
  \hspace*{1cm}
  \subfloat{{\includegraphics[width=0.4\linewidth]{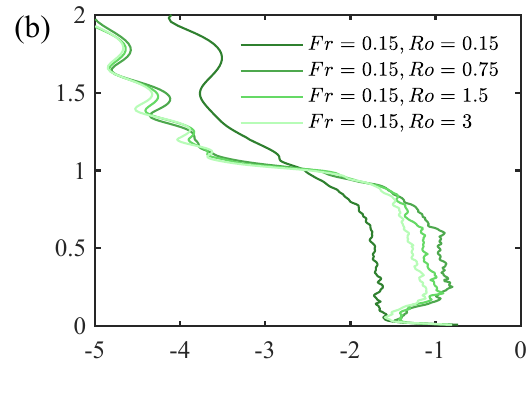}}}%
  \\ 
  \subfloat{{\includegraphics[width=0.4\linewidth]{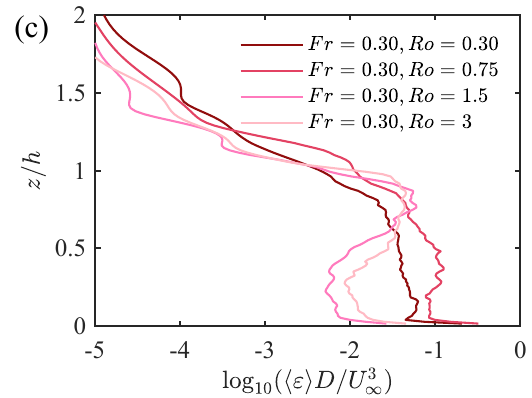}}}%
  \hspace*{1cm}
  \subfloat{{\includegraphics[width=0.4\linewidth]{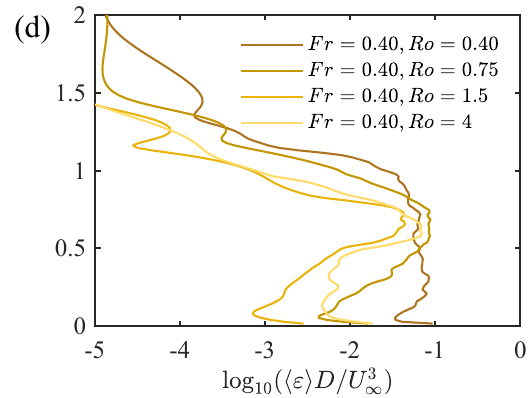}}}%
  \caption{TKE dissipation $\langle \varepsilon \rangle$ measured at $y=0,\, x/D=1$. (a) $Fr=0.075$, (b) $Fr=0.15$, (c) $Fr=0.30$, and (d) $Fr=0.40$. } \label{dissp_y0_is} 
\end{figure*}

Figure \ref{dissp_x1_y} shows the vertical profile of instantaneous (at the same instances as in Fig.~\ref{dissp_rig}(a-d)) and time-averaged TKE dissipation at $x/D=1, y=0$. As stratification increases, the vertical length scale of
the instantaneous dissipation patches (approximately $\propto U_{\infty}/N$) decreases, the spikes in their profiles
increase in sharpness, and the number of spikes also increases.
Quantitatively, the magnitude of time-averaged dissipation in the wake reaches $\langle \varepsilon \rangle \sim  10^{-1} \,U_{\infty}^3/D$, while the instantaneous peak value could be an order of magnitude higher, as shown in Fig.~\ref{dissp_x1_y} (a,b). Here, dissipation is represented in the inertial units, $U_{\infty}^3/D$, which can be scaled up or down for varying current speeds and seamount dimensions, and is equivalent to $2 \times 10^{-6}\, {\rm W\, kg^{-1}}$ for $U_{\infty}=0.1 \, {\rm m\, s^{-1}}$ and $D = 500 \, {\rm m}$. The magnitude $O(10^{-1}) \,U_{\infty}^3/D$ is consistent in different cases, while the dissipation {in the ambient} is around $10^{-5}\sim 10^{-4}$ ($U_{\infty}^3/D$), more than 1000 times lower. 
Appendix \hyperref[re_sens]{B} shows that at the present Reynolds numbers, the dissipation is relatively independent of $Re$.

\begin{figure*}[thb]
  \centering 
  \captionsetup[subfloat]{farskip=-4pt,captionskip=1pt}
  \subfloat{{\includegraphics[width=.5\linewidth]{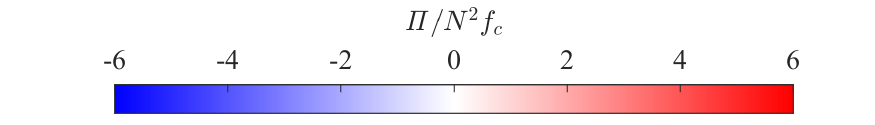}}} \hfill \\
  \subfloat{{\includegraphics[width=0.5\linewidth]{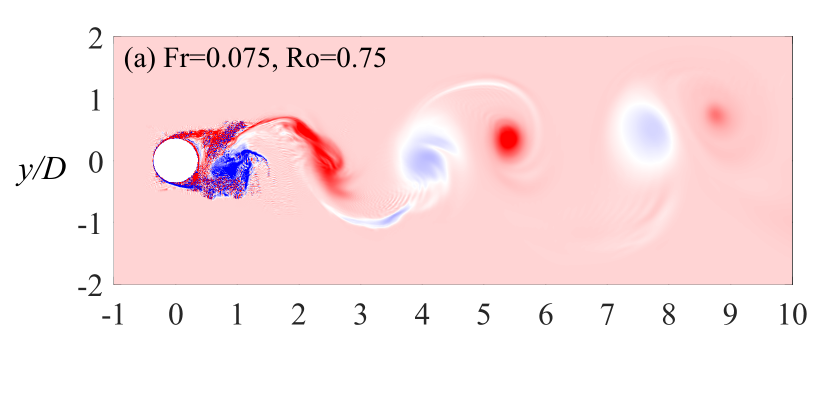}}}%
  \subfloat{{\includegraphics[width=0.5\linewidth]{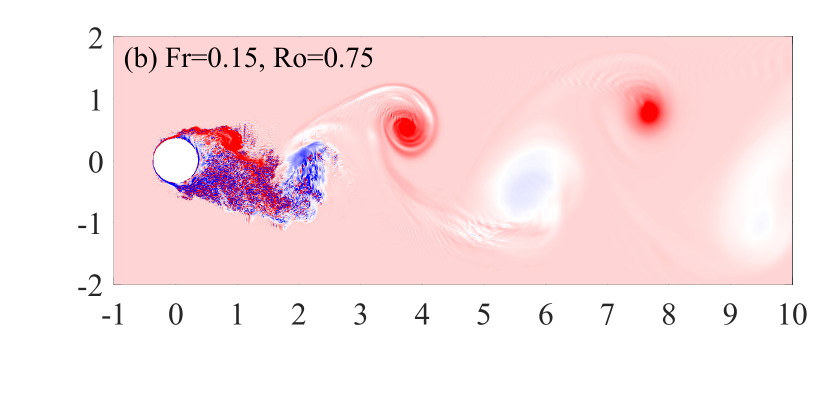}}}%
  \\ 
  \subfloat{{\includegraphics[width=0.5\linewidth]{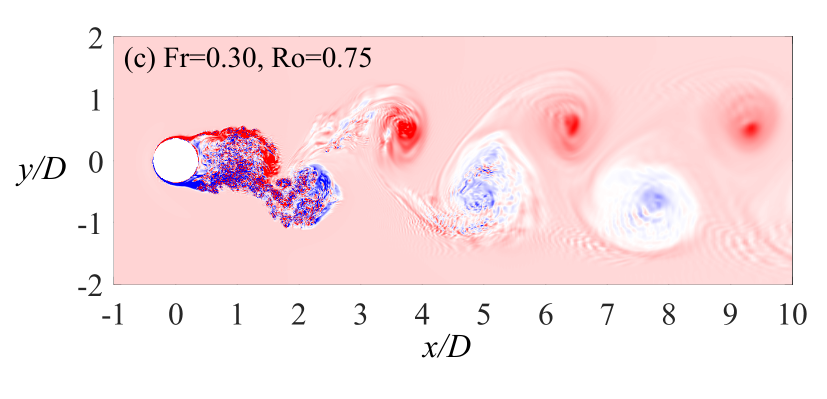}}}%
  \subfloat{{\includegraphics[width=0.5\linewidth]{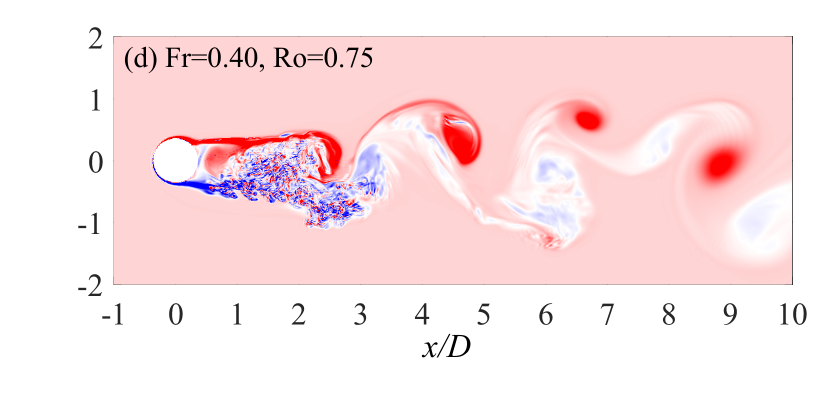}}}%
  \caption{Contours of the  {normalized} potential vorticity, ${\it \Pi}/N^2 f_c$, in the horizontal plane at $z/h=0.25$.  {Note that $f_c$ is signed and $f_c<0$ in the present work, such that negative normalized PV resides on the $y<0$ side, in general.} Panels (a-d): $Fr=0.075, 0.15, 0.30, 0.40$, respectively, and $Ro=0.75$ for all panels. The background value is unity for all panels, corresponding to normalized background PV (unity) due to the system rotation. } \label{pv_z_025}
\end{figure*}

The dissipation obtained in the present simulations compares qualitatively and quantitatively well with observational data, in the wakes of Palau \citep{mackinnon2019eddy,st2019turbulence,wijesekera2020observations,wynne2022measurements}, the Green Island \citep{chang2013kuroshio}, and seamounts in the Tokara Strait \citep{nagai2021kuroshio}. Spatially localized dissipation sites and spiky vertical dissipation profiles were also observed in \cite{mackinnon2019eddy,wynne2022measurements}.  The {\it in situ} measured dissipation magnitude lies between $10^{-7}\sim 10^{-5}\,{\rm W\, kg^{-1}}$ in \cite{chang2013kuroshio},  $10^{-6}\sim 10^{-4}\,{\rm W\, kg^{-1}}$ in \cite{wijesekera2020observations}, and $10^{-7}\sim 10^{-6}\,{\rm W\, kg^{-1}}$ in \cite{nagai2021kuroshio}, all in general agreement with the present results. This range of dissipation values in the observations could be due to a number of factors, for example, differences in topography size, in strengths of the mean current and tidal flows, and in the background $N$, but an overall agreement within measurements and between measurements and the present simulations is reached. It is also noted that the tidal component, which is another destabilizing factor, has not been included here and is the subject of separate study for situations where its 
magnitude is as strong as or even stronger than the current.









Figure  \ref{dissp_y0_is}  compares the dissipation profiles at $x/D=1,y=0$ for different $(Fr,Ro)$ cases. The dual role of rotation can be seen. On one hand, as shown in Fig.~\ref{dissp_y0_is}(c),  case Fr030Ro075 
  {with an intermediate rotation rate} has the largest dissipation at the centerline  {compared to other} 
 $Ro$ values that are higher or lower. 
 {This is also confirmed later to be the case in volume-integrated $\varepsilon$ and shown to be linked to CI.}
On the other hand, in Fig.~\ref{dissp_y0_is}(a-b), very strong rotation ($Ro=0.15$) significantly reduces $\varepsilon$ by almost an order of magnitude compared to $Ro \ge O(1)$, through the reduction of the vertical velocity gradients and the consequent  vertical shear instability. Moving from Fig.~\ref{dissp_y0_is}(b) to (c), the effect of rotation appears to be non-monotonic due to the reasons mentioned above, but a more comprehensive measure of the dissipation in the 3D domain is required and the examination of the CI is needed. These investigations are reserved for the next sections.

 {It is shown in Fig.~\ref{dissp_rig}(g,h) that the vertical shear that leads to KHI can originate from the vortex structures in the wake and the HJ near the top of the seamount. These two mechanisms are distinct and both play an important role in the dynamics of topographic flows. A quantitative assessment reveals their relative contribution. In the center plane ($y=0$), the HJ region (appropriately $1-1.2Fr<z/h<1$) contributes 24-32\% and 35-51\% of the planar dissipation in cases Fr030 and Fr040, respectively. Note that this range of $1-1.2Fr<z/h<1$ is nearly half of the wake height ($0<z/h<1$) in case Fr040 so the large percentage is not surprising. In volumetric measures (which are analyzed in detail in section \ref{param}), the similarly defined HJ region  contributes less than 3\% to the total volume-integrated dissipation in cases Fr030 and Fr040 and less at lower $Fr$. The implication is two-fold. On one hand, the significance of HJ to dissipation cannot be ignored, especially in the vertical center plane -- where it is the strongest. On the other hand, in flows around 3D topographies, its volumetric contribution is much smaller compared to the wake core, unlike in 2D topographic flows. } 


\subsection{Contribution of the centrifugal/inertial instability (CI)}

Planetary rotation substantially changes the spatial organization of flow structures that are large enough to feel it ($Ro<O(10)$), and alters the dissipation in various ways, both direct and indirect. The direct effect of rotation can be sensed by  CI, whose intensity has a parametric dependence on rotation and is stronger at $Ro=O(1)$ (see Appendix \hyperref[theory]{C}). The indirect effect, although still associated with CI, comes from the change of the mean/base flow through the modification of the coherent structures. For the latter, eddies that rotate in the same direction as the system rotation (cyclonic vortices, CVs) and those rotate in the opposite direction (anticyclonic vortices, AVs) behave differently. 
{Among  flows with $| \omega_z | = O(f)$}, anticyclonic vortices and shear are often subject to CI, 
 leading to an appreciable asymmetry between the two sides of the wake. We note that in the present wake, planetary rotation is clockwise as in the Southern Hemisphere and AVs are primarily on the right-hand side ($y<0$ and $\omega_z >0$).  

\begin{figure*}[t!]
  \centering 
  \captionsetup[subfloat]{farskip=0pt,captionskip=1pt}
  \subfloat{{\includegraphics[width=0.3\linewidth]{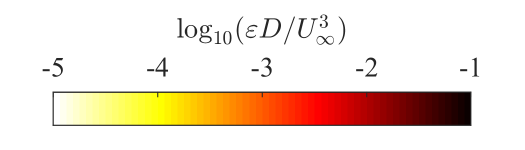}}}%
  \\ 
  \subfloat{{\includegraphics[width=0.25\linewidth]{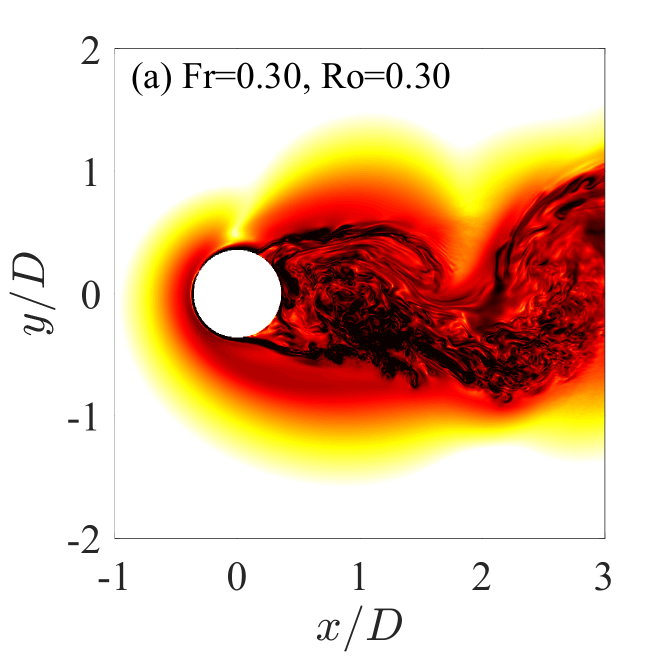}}}%
  \subfloat{{\includegraphics[width=0.25\linewidth]{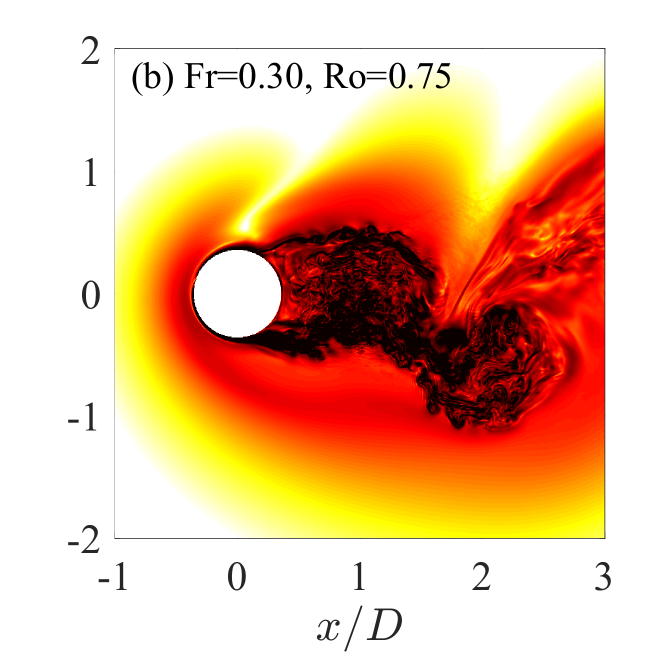}}}%
  \subfloat{{\includegraphics[width=0.25\linewidth]{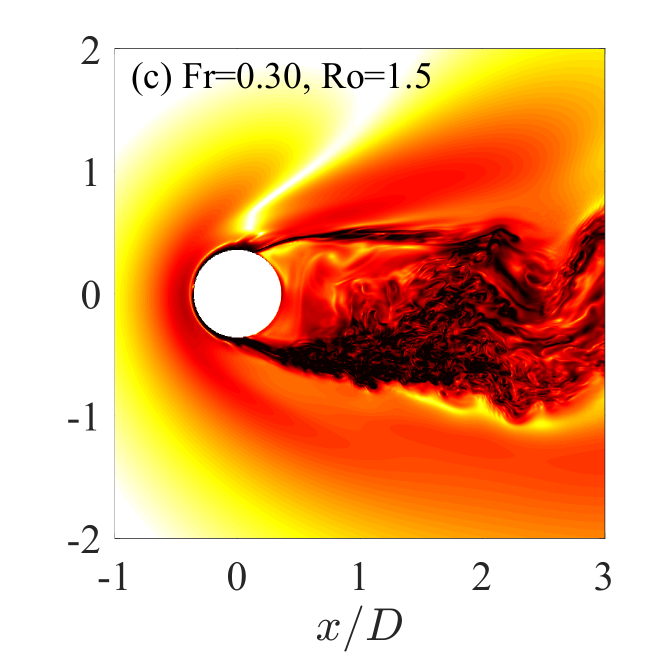}}}%
  \subfloat{{\includegraphics[width=0.25\linewidth]{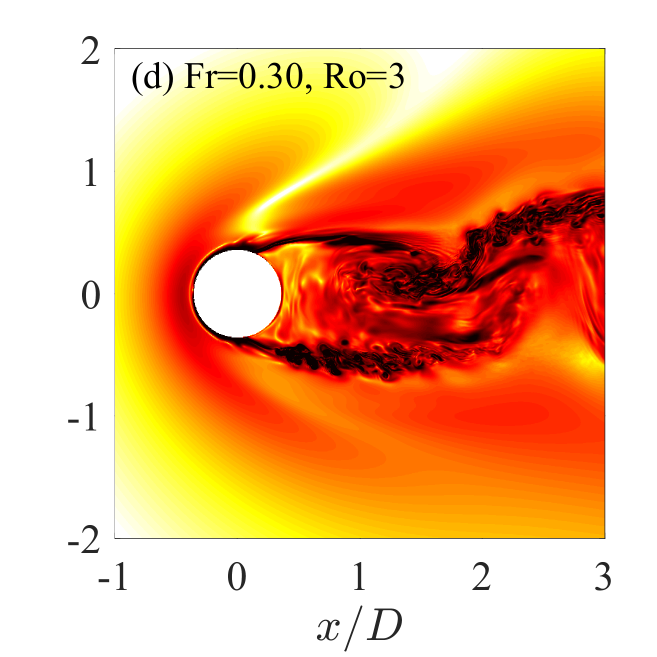}}} \\
  \subfloat{{\includegraphics[width=0.33\linewidth]{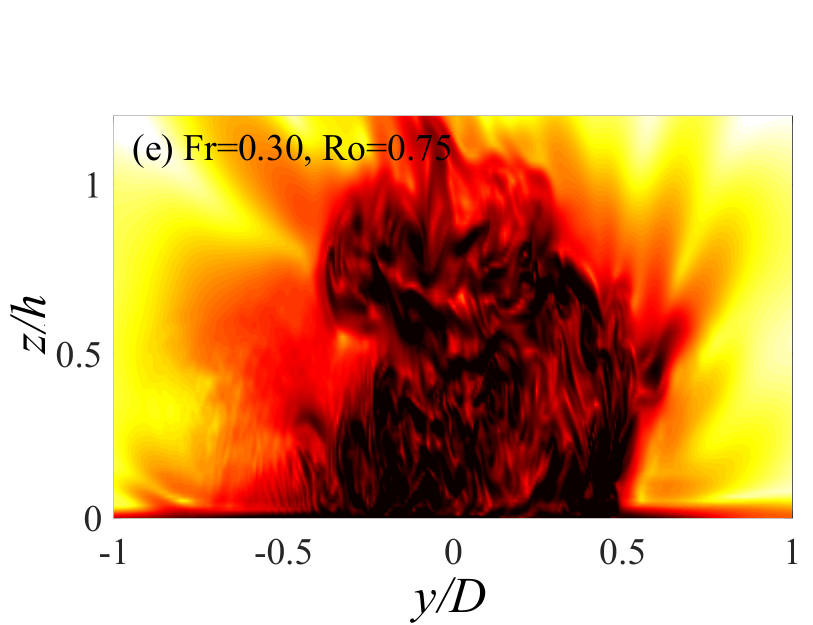}}} 
  \subfloat{{\includegraphics[width=0.33\linewidth]{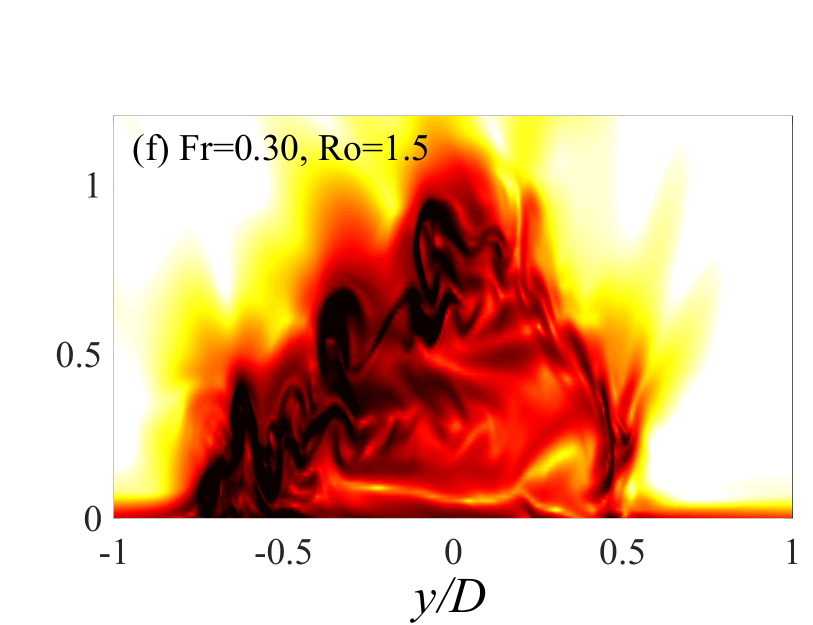}}} 
  \subfloat{{\includegraphics[width=0.33\linewidth]{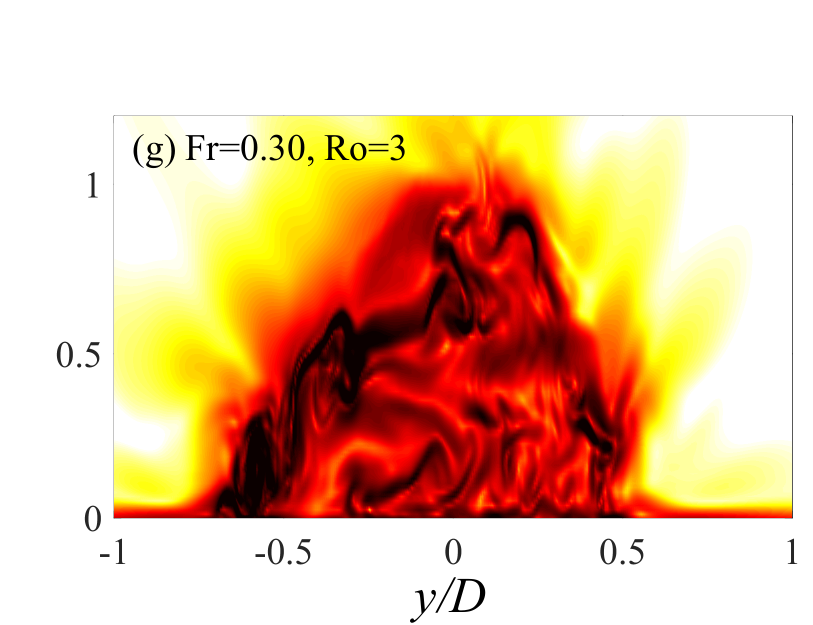}}} 
  \caption{Instantaneous dissipation rate $\varepsilon$ in (a-d) the $z/h=0.25$ plane and (e-g) the $x/D=1$ plane. 
  Panels (b,e), (c,f), and (d,g) correspond to the same time instances, respectively.} \label{dissp_z_025} 
\end{figure*}

We move on to the potential vorticity (PV),  
 \begin{equation}
  {\it \Pi} = (\bm{f}_{c}  + \bm{\omega}) \cdot \nabla \tilde{b},   \label{pv}
\end{equation} 
where $\bm{\omega }= \nabla \times \bm{u}$ is the relative vorticity and $\tilde{b}=b+N^2z$ is the `total buoyancy'. PV is a useful diagnostic in  large-scale flow analyses since it is conserved along isopycnal surfaces in the absence of friction or mixing. 
On the other hand, the sign of PV 
 has implications for scales smaller than the balanced motions.  {PV of opposite sign to $f_c$}  serves as an indicator of several hydrodynamic instabilities \citep{thomas2013symmetric}, that are precursors to turbulence. The horizontal and vertical components of PV,  \begin{align}
  \it{\Pi}_h &= \omega_x \frac{\partial \tilde{b}}{\partial x} + \omega_y \frac{\partial \tilde{b}}{\partial y}; \;
  \it{\Pi}_v = (f_{ c} + \omega_z) \frac{\partial \tilde{b}}{\partial z}, 
\end{align} are indicative of the symmetric and centrifugal/inertial instabilities, respectively \citep{thomas2013symmetric}. 

\begin{figure*}[t!]
  \centering 
  \captionsetup[subfloat]{farskip=0pt,captionskip=1pt}
  \subfloat{{\includegraphics[width=0.25\linewidth]{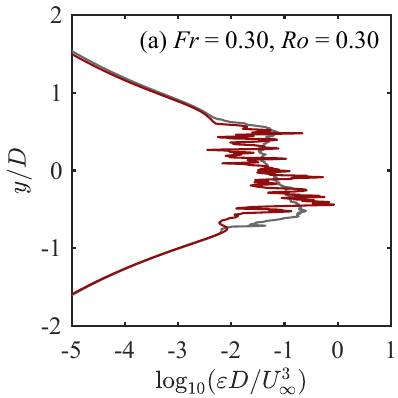}}}%
  \subfloat{{\includegraphics[width=0.25\linewidth]{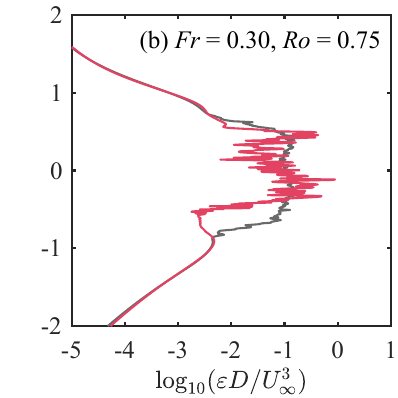}}}%
  \subfloat{{\includegraphics[width=0.25\linewidth]{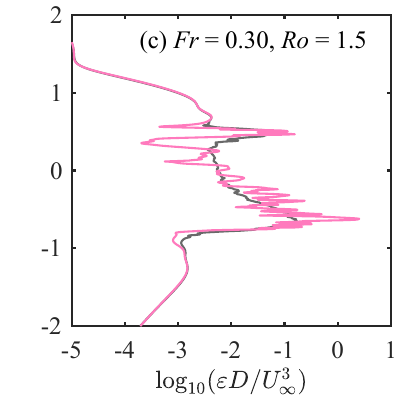}}}%
  \subfloat{{\includegraphics[width=0.25\linewidth]{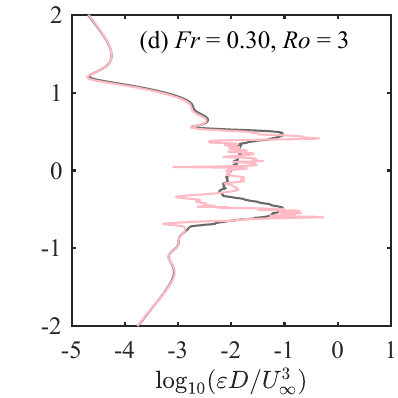}}}%
  \caption{Instantaneous dissipation rate $\varepsilon$ (colored) and its time-average $\langle {\varepsilon} \rangle$ (TKE dissipation, gray), probed on the horizontal line $z/h=0.25, x/D=1$. The time instances are the same as those in Fig.~\ref{dissp_z_025}(a-d). 
  } \label{dissp_x1_z}
\end{figure*}


Figure \ref{pv_z_025} shows  PV on the horizontal  plane $z/h=0.25$ at $Ro=0.75$ and different $Fr$. In the near wake of each case, entangled fine-scale structures of positive and negative  {(normalized)} PV can be seen, as a result of near wake turbulence.  As the flow evolves into the intermediate wake, large patches of negative PV (indicated by blue color) can be seen on the anticyclonic side ($y<0$), with the magnitude decaying in the streamwise direction as the flow gradually adjusts to rotation, eventually reaching a near-zero PV, stable state with little turbulence. On the cyclonic side ($y>0$), strong large-scale coherent CVs can be found with positive PV, and they remain intact from destruction during their downstream propagation.  {This asymmetry between anticyclonic and cyclonic vorticity is a characteristic feature of submesoscale rotating flows with $Ro = O(1)$ and it will be analyzed quantitatively in section} \ref{param}.  

At the $Fr \leq O(1)$ values of this case study, the flow exhibits a von K{\'a}rm{\'a}n shedding pattern in horizontal planes.    Additionally, the case-dependent $Fr$ introduces a subtle difference in the vortex dynamics. At the lowest $Fr$ ($Fr=0.075$, Fig.~\ref{pv_z_025}(a)), dipoles are formed due to strong mutual interaction between the AVs and the CVs, where the CVs are systematically stronger and they attract AVs to the cyclonic side, leading to the veering of the wake. The increased horizontal vortex--vortex interaction that is closer to 2D dynamics is a consequence of strong stratification and limited vertical motions. This behavior is similar to the Bu25 case in \cite{liu2024effect}, where the vortices were tracked in time and  dipole formation was statistically shown by the mean vortex trajectories and conditional vorticity distribution. 
At $Fr=0.15,0.30$ (Fig.~\ref{pv_z_025}(b-c)), the overall shedding pattern mimics that of a standard K{\'a}rm{\'a}n street while the CI of the AVs is more appreciable visually. At $Fr=0.40$ (Fig.~\ref{pv_z_025}(d)), the recirculation zone is significantly longer than that in other cases (which can also be seen in Fig.~\ref{dissp_rig}) and the vortex shedding pattern is less regular. The former is due to the fact that the hydraulic jet reaches a lower downward distance at  {higher $Fr$}, and it interacts strongly with the separation \citep{chomaz1993structure}. The vertical PV (${\it \Pi}_v$, not shown) is similar to the total PV, and the horizontal PV (${\it \Pi}_h$, not shown) did not show evidence of the symmetric instability. 






Figure \ref{dissp_z_025} shows instantaneous $\varepsilon$ at $Fr=0.30$ and various $Ro$, where both the K{\'a}rm{\'a}n vortices and the shear layers present instabilities due to rotation. 
In all cases, small-scale turbulence structures can be seen in the near wake, whose spatial location coincides well with those wake eddies and the shear layer. These worm-like structures mimic the intense, randomly oriented vortex tubes in isotropic turbulence and are indicative of fully triggered turbulence as a consequence of the breakdown of the 3D instabilities and the establishment of a forward cascade. However, wake turbulence decays in the streamwise direction due to the stabilizing effect of strong stratification. Hence, these fine-scale structures do not  persist long after their generation and turbulence is localized to the near wake, which can also be seen in Fig.~\ref{pv_z_025}. {The time-evolution of $\varepsilon$ and $\varepsilon_{\rho}$ for panels (a-d) can be found in the supplementary materials (movies). The spatial structure of $\varepsilon_{\rho}$ is similar to $\varepsilon$.}

\begin{figure*}[h!]
  \centering 
  \captionsetup[subfloat]{farskip=0pt,captionskip=1pt}
  \subfloat{{\includegraphics[width=0.475 \linewidth]{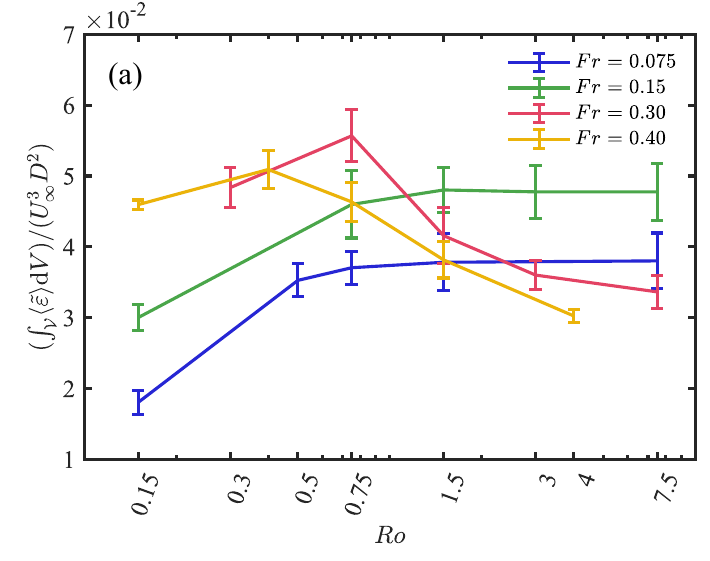}}} 
  \hspace{0.5cm}
  \subfloat{{\includegraphics[width=0.475 \linewidth]{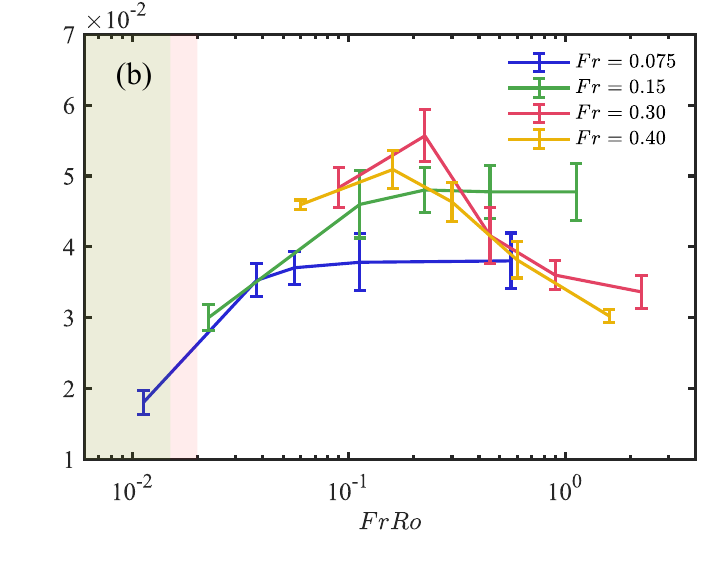}}}%
  \\
  \vspace{-0.2cm}
  \subfloat{{\includegraphics[width=0.475 \linewidth]{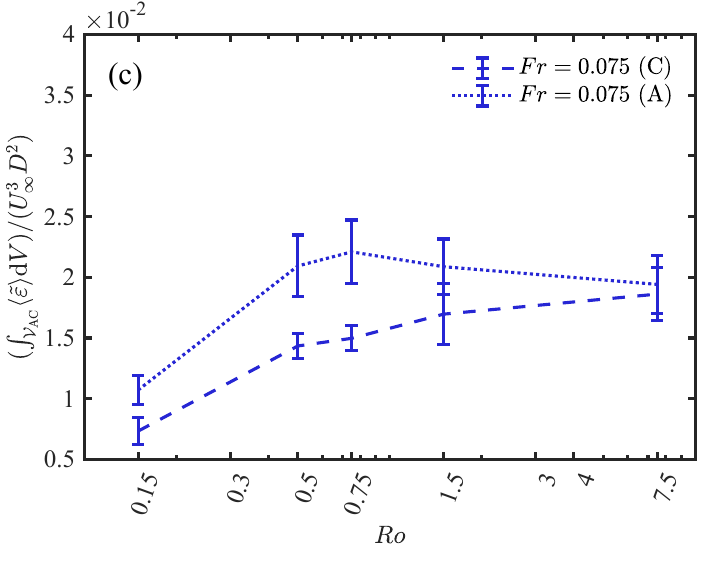}}} 
  \hspace{0.5cm}
  \subfloat{{\includegraphics[width=0.475 \linewidth]{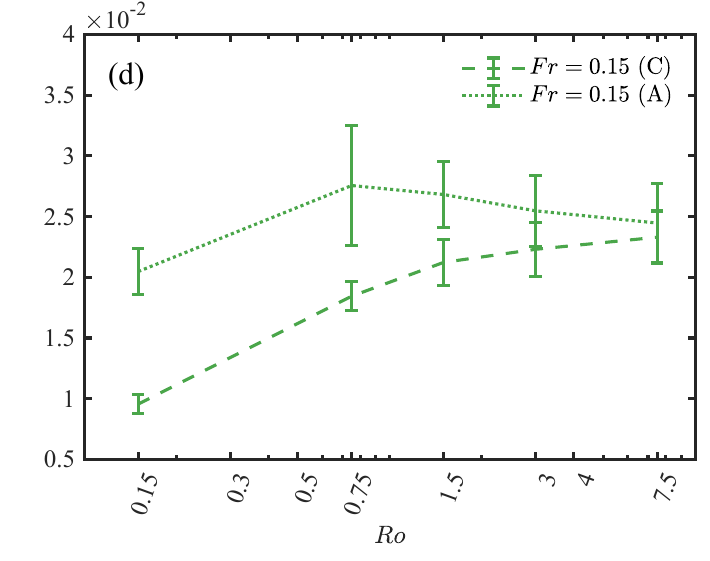}}}%
  \\
  \vspace{-0.2cm}
  \subfloat{{\includegraphics[width=0.475 \linewidth]{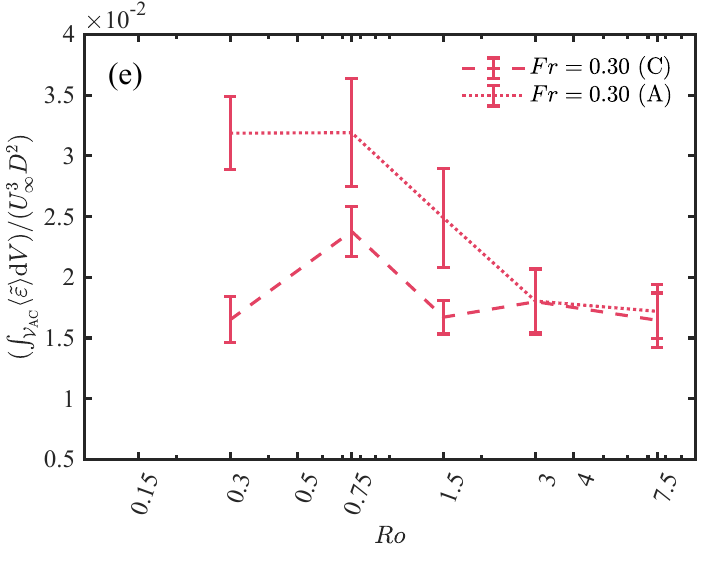}}} 
  \hspace{0.5cm} 
  \subfloat{{\includegraphics[width=0.475 \linewidth]{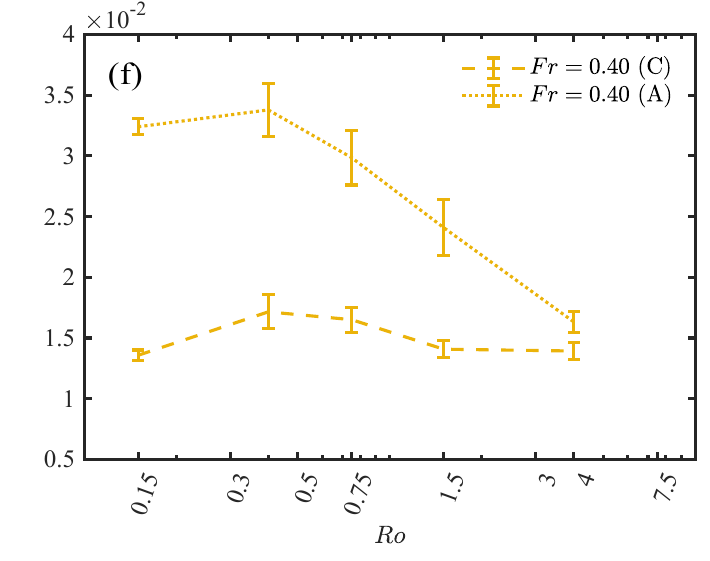}}}%
  \caption{Volume-integrated dissipation, $\int_{\mathcal{V}} \langle \tilde{\varepsilon} \rangle \,{\rm d}V $,  as a function of (a) $Ro$ and (b) $FrRo$. Here the integration domain is $\mathcal{V}=[-D, 4D]\times[-2D,2D]\times[0,2h]$. The vertical bars denote one standard variance above and below the averages. The green and red shades in (b) mark the overlap of the present parameters and those in \cite{perfect2020energetics1} and \cite{srinivasan2021high}, respectively.  {(c-f): similar to (a), but the volume $\mathcal{V}_{\rm AC}$ contains only the cyclonic ($y>0$, marked with `C') or the anticyclonic ($y<0$, marked with `A') side.} } \label{dissp_para} 
\end{figure*}

At $Fr=0.30$, case Fr030Ro075 show the strongest turbulent dissipation (also evidenced by volumetric dissipation shown later in Fig.~\ref{dissp_para}), with the turbulent worm patches fully filling the interior of the eddies and crossing the centerline. Case Fr030Ro030 and Fr030Ro1p5 both have less dissipation, but they present the strongest lateral asymmetry, with the anticyclonic side unstable to CI that generates turbulence and the cyclonic side being much more stable. In case Fr030Ro3, the Burger number is $Bu=100$ and the effect of rotation is relatively small. It shows  weaker dissipation but both the AV and CV sides display braids of instability structures.  

The aforementioned dynamics are also reflected in the statistics, shown in Fig.~\ref{dissp_x1_z}. In cases Fr030Ro030 and Fr030Ro1p5, the asymmetry between AV and CV sides is evident, with the AV dissipation having a wider spread and a higher magnitude. The dissipation asymmetry is the largest in the Fr030Ro1p5 case, where the anticyclonic peak $\varepsilon$ is almost an order of magnitude higher than the cyclonic $\varepsilon$. In case Fr030Ro075, the asymmetry is less appreciable as dissipation fills more space near the centerline than in other cases. In the Fr030Ro3 case, the symmetry of the AV and CV dissipation gets close to the limit of a non-rotating case. 
The above scenarios are also qualitatively and quantitatively similar in the $Fr=0.40$ cases with similar $Ro$ (not shown).

Now we turn our attention to $y$-$z$ planes in the near wake ($x/D=1$), as shown in Fig.~\ref{dissp_z_025}(e-g), in which the structures of CI-induced dissipation can be seen more clearly. For cases Fr030Ro1p5 and Fr030Ro3 in (b,c),  rolled up dissipation structures  can be seen at various heights on the anticyclonic (left) side, while the cyclonic shear layer is more stable (at this $x/D$ location). These  dissipation structures are associated with the quasi-streamwise vortices during the growth period of the CI, and were also found in other rotating flows, e.g., in \cite{kloosterziel2007inertial,arobone2012evolution,carnevale2013inertial}. 
{A linear analysis of the growth of the streamwise/azimuthal vorticity during CI is provided in Appendix \hyperref[theory]{C}, which demonstrates its relevance to the diagnosis of CI and the relation between the growth rates and respective instability criteria. }









In Fig.~\ref{dissp_z_025}(e), the wake of case Fr030Ro075 is filled with three-dimensional turbulent vortices (`worms'), unlike the wakes in (f,g), which have a hollow core enclosed by the shear layers. These `worms' are the results of the subsequent nonlinear interaction and rapid three-dimensionalization after the onset of CI, and the strong interaction of the anticyclonic and cyclonic vortices/shear layers. The more intense structures and more `mature' turbulence in this case are both suggestive of stronger CI. It is also noted that CI is baroclinic/three-dimensional and, similar to KHI,  it also distorts the isopycnals and enhances mixing. 

\section{{Parametric dependence of dissipation on stratification and rotation}} \label{param}




\subsection{ {Volumetric measures}} 

 KHI and CI, which were discussed separately, in the previous section
co-exist and their combined influence is responsible for the dependence of  wake turbulence on stratification and rotation. An overall measure of turbulence is chosen  for quantifying the dependence. Specifically, the time- and volume-integrated dissipation
  \begin{equation}
  \mathcal{E}  = \int_{\mathcal{V}} \langle \tilde{\varepsilon}  \rangle \, {\rm d}V \label{dissp_int}
\end{equation} 
is employed. The size of the integration subdomain is $\mathcal{V}=[-D,4D]\times[-2D,2D]\times[0,2h]$ that encloses the turbulent near wake. Here $\langle \tilde{\varepsilon}  \rangle$ is the time-averaged (over $N_t \approx 30$ for 3D snapshots) dissipation rate of the total kinetic energy (instead of TKE), for better converged statistics and greater relevance to field measurements.  






Figure \ref{dissp_para}(a,b) show the time- and volume-integrated dissipation rate $\mathcal{E}$ for each $Fr$-series as a function of  $Ro$ in (a)  and  $FrRo$ in (b).   
It can be seen that the data break into two groups: (1) the lower-$Fr$ group, Fr007 and Fr015, and (2) the higher-$Fr$ group, Fr030 and Fr040. 

In the first group with very strong stratification, the effect of rotation appears to be solely stabilizing. The dissipation increases as $Ro$ increases, and eventually saturates at high $Ro$ or  weak rotation. The dissipation in Fr007 cases is consistently lower than that in Fr015 cases, even though the $Re$ is  {twice as high as} in the Fr007 cases.
The trends of monotonic decrease of dissipation with decreasing $Ro$ and $Fr$ are qualitatively consistent with the findings and the parameter ranges in previous ROMS studies of \cite{perfect2020energetics1} and \cite{srinivasan2021high}. A difference is that the weak rotation (higher $Ro$) saturation was not observed in the previous work since it concerned the  strong rotation/large topography regime of $Fr,Ro \le O(0.1)$. In Fig.~\ref{dissp_para}(b), the green and red shaded regions  mark the upper end of the parameter combinations of $FrRo$ in \cite{perfect2020energetics1} and \cite{srinivasan2021high}. 


In the second group where stratification is not too strong, the effect of rotation is non-monotonic -- there is an intermediate value of $Ro$ where the volume-integrated dissipation peaks. The intermediate value is $Ro=0.40$ for the Fr040 cases and $Ro=0.75$ for the Fr030 cases. Both peaks fall within $Ro=O(0.5-1)$, corresponding to SMS topographies or eddies. The dissipation maxima at the SMS and the associated most destabilizing rotation have not been well explored in the previous parameterizations of topographic wakes and highlight the significance of CI and its $Ro$-dependence.

Similar non-monotonic rotation dependence was seen in the rotating horizontal shear layer with and without vertical stratification \citep{yanase1993rotating,arobone2012evolution}, and is  characteristic of CI. The numerical stability analysis of \cite{yanase1993rotating} and \cite{arobone2012evolution} revealed that, with no rotation, the three-dimensionally most unstable mode is the 2D KH mode ($k_z=0$), which is still the case when there is vertical stratification \citep{arobone2012evolution}.  {As the authors found}, when there is weak rotation  ($Ro=O(10)$), the most unstable mode is still the KH mode, but a nearly streamwise invariant inertial mode starts to emerge. As the rotation rate increases and $Ro$ reaches $O(1)$, the growth rate of the inertial instability overtakes that of the KH mode. However, when rotation is further increased to $Ro \sim O(0.1)$, the inertial mode disappears. The existence of a most destabilizing rotation rate can be shown in a linear analysis of both parallel and circular flows in Appendix \hyperref[theory]{C}, which explains the existence of  {a dissipation peak at $Ro=O(0.5)$} in the Fr030 and Fr040 cases when stratification is not too strong.   {Cases Fr007 and Fr015 do not have this behavior.} 

 {Unlike studies focused solely on CI \citep{yanase1993rotating,kloosterziel2007inertial,arobone2012evolution,carnevale2013inertial} where cyclonic and anticyclone shear/vortices are treated separately, the present wakes contain both, leading to anticipated asymmetry. Thus, the volumetric measure in Fig.~\ref{dissp_para}(a) is repeated in (c-f) but with the cyclonic and the anticyclonic sides considered separately. 
 {Overall, dissipation asymmetry decreases with weakening rotation, approaching symmetry at $Ro\ge O(5)$ across a range of $Fr$.} 

In cases Fr030 and Fr040 (Fig.~\ref{dissp_para}(e,f)), it is clear that the anticyclone side is strongly influenced by rotation, and that the dissipation increases rapidly as rotation strengthens until it saturates in the $Ro\le O(0.5)$ range. On the other hand, the cyclonic side is rather weakly influenced by rotation except for cases Fr030Ro075 and Fr040Ro040. The cyclonic dissipation peaks are  presumably due to the instability of the cyclonic shear layer/vortices and strong interactions between cyclonic and anticyclonic sides (see Fig \ref{dissp_z_025}(b) and the supplementary movies). It is also noted in Appendix  \hyperref[theory]{C} that cyclonic vortices can also become unstable to CI.    

In  {the strongly stratified} cases Fr007 and Fr015 (Fig.~\ref{dissp_para}(c,d)),  dissipation on the cyclonic side increases monotonically with $Ro$, suggesting that the cyclonic side is subject to the stabilizing  effect of rotation. Anticyclonic dissipation, however, displays a mild peak at $Ro=0.75$ for both cases, suggesting that CI is attenuated but still leaves a signature in the dissipation. Intuitively, as stratification becomes stronger, such as in cases Fr007 and Fr015, the vertical length scale for instability decreases and the space for CI shrinks until it is eventually suppressed.  {The anticyclonic CI attenuation and the cyclonic stabilization add up to a net effect of rotation being solely stabilizing at $Fr=0.075, 0.15$.}}




{The  present parametric dependence has several implications. First, the  inertial  scaling $\varepsilon = C_{\varepsilon} U^3_{\infty}/D$, where $C_{\varepsilon}$ is a scaling {coefficient}, works reasonably well for the wake dissipation over a wide range in the parameter space of $(Fr,Ro)$ (the value of $FrRo$ varies by 2 orders of magnitude). It provides a means for contextualizing field measurements or modeling wake dissipation/mixing in regional or global climate models. On the other hand, the variability of $C_{\varepsilon}$ over the range of $Fr$ and $Ro$ investigated reflects systematic effects of stratification and rotation. It is evident that turbulence in seamount or hill wakes at different levels of stratification depends differently on rotation, with a transition point  between $Fr=0.15$ and $Fr=0.30$ due to the activation of the CI. That being said, $C_{\varepsilon}$ is likely not a simple function of $Fr$ and $Ro$.}

In previous numerical studies of topographic wakes \citep{perfect2018vortex,perfect2020energetics1,srinivasan2021high}, it was suggested that the vortex dynamics and turbulent dissipation might both be categorized by a single parameter, instead of $(Fr,Ro)$. For example, \cite{perfect2018vortex} varied both $Fr$ and $Ro$ in the range of $0.014 \le Fr^* \le 0.14 $ and $0.053 \le Ro^* \le 0.21$, which is equivalent to $0.014 \le Fr \le 0.14 $ and $0.0265 \le Ro \le 0.105$ after a conversion to our definition ($Fr=Fr^*, Ro=Ro^*/2$).  It was suggested that the vortex dynamics can be characterized as a function of the Burger number, $Bu=(Ro/Fr)^2=(Nh/f_cD)^2$, and the dissipation can further be parameterized as a function of a positive power of $FrRo$. \cite{srinivasan2021high} studied topographic wakes as a function of $Ro$ ($0.025 \le Ro \le 1$) when $Fr=0.02$ is fixed, and found that the dissipation monotonically increases as a function of the Rossby number. For comparison, the same data in Fig.~\ref{dissp_para}(a) is plotted in Fig.~\ref{dissp_para}(b), as a function of $FrRo$. Although the data seems to collapse better, the division into two groups with different dependencies is still very clear. The parameters in \cite{perfect2020energetics1} have $(FrRo)_{\rm max} \approx 0.015$ and those in \cite{srinivasan2021high} have $(FrRo)_{\rm max}=0.02$. Both overlap with the lower end of the present parameters, as shown by the green and red shades in Fig.~\ref{dissp_para}(b).




It was pointed out by \cite{liu2024effect} that there is still vertical coupling of the vortex shedding at $Fr=0.15, \, Bu=\infty$ and hence the $Bu$-determination of coupling/decoupling is incomplete without an additional $Fr$-dependence -- stronger stratification than $Fr=0.15$ is required to vertically decouple the vortex shedding in rotating and non-rotating K{\'a}rm{\'a}n wakes. A similar role of stratification is found in the dissipation dependence,  {that the activation or attenuation of the CI between $Fr=0.15$ and $Fr=0.30$ changes the $Ro$-dependence of dissipation. Thus, $Fr$  plays a crucial role as a separate independent parameter in the regime transitions of vortex dynamics and turbulence generation. It requires separate  attention instead of being simply absorbed into $Bu$ or $FrRo$. Last but not least, the asymmetry between the cyclonic and anticyclonic sides is significant at moderate to strong rotation. The difference between the responses of cyclonic and anticyclonic shear layers/vortices to changing $Ro$ requires separate consideration in the analysis and parameterization of topographic wakes.}

\begin{figure*}[t!]
  \centering 
  \captionsetup[subfloat]{farskip=-6pt,captionskip=1pt}
  \subfloat{{\includegraphics[width=0.45\linewidth]{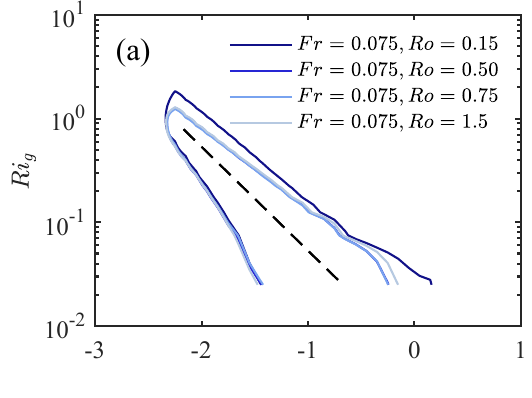}}}%
  \hspace*{1cm}
  \subfloat{{\includegraphics[width=0.45\linewidth]{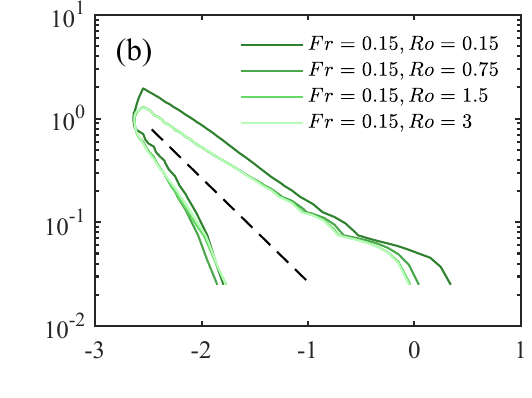}}}%
  \\ 
  \subfloat{{\includegraphics[width=0.45\linewidth]{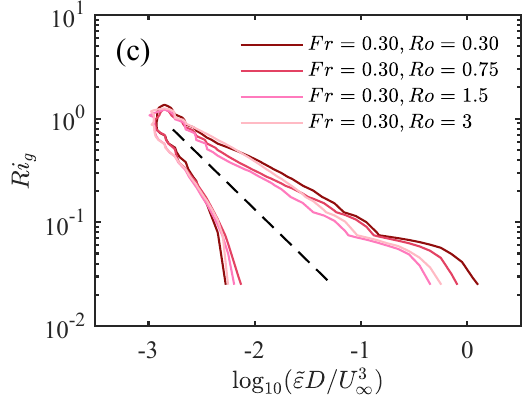}}}%
  \hspace*{1cm}
  \subfloat{{\includegraphics[width=0.45\linewidth]{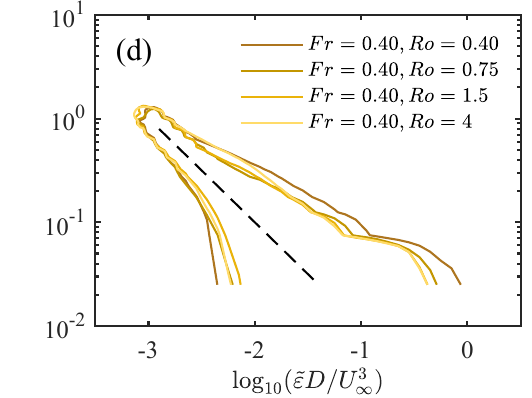}}}%
  \caption{Joint p.d.f (JPDF) of the total KE dissipation ($\tilde{\varepsilon}$, defined in \eqref{eps_def}) and the local gradient Richardson number ($Ri_g$, defined in \eqref{rig_def}).  Contours enclose $85\%$ of the JPDF in each case. Plane location $z/h=0.25$. The $Fr$-dependent black dashed lines in each figure are given by $\tilde{\varepsilon} D/U_{\infty}^3 = C_R Fr^{-1} Ri_g^{-1}$, with the same fitting constant $C_R = 1/2500$.  
  } \label{jpdf_dissp_rig}  
\end{figure*}

\subsection{ {Relating dissipation rate to local stability criteria}}

 {Besides global measures such as volume-integrated dissipation as a function of $Fr$ and $Ro$, local measures such as the joint distributions of stability criteria and dissipation could provide additional information for parameterization. Possible surrogates include the gradient Richardson number for KHI and PV for CI and the overarching goal is to seek whether,  given  information about the local stability criteria, the prediction of dissipation can be improved. }

Figure \ref{jpdf_dissp_rig} shows the joint probability distribution functions (JPDFs) of the instantaneous total KE dissipation ($\tilde{\varepsilon}$) and the local gradient Richardson number ($Ri_g$).  {Here $\tilde{\varepsilon}$ and $Ri_g$ are both based on instantaneous velocity components/buoyancy as previously defined in \eqref{eps_def} and \eqref{rig_def}.}    {The shear-stable background flow with $\tilde{Re}_b<0.1,Ri_g>O(1)$ is excluded from the JPDF, allowing a focus on the shear-unstable states. The contours in each plot enclose the most probable 85\% of the data/ensemble.
When this percentage is changed to 80\% and 90\%, similar results are obtained.} The shape of the JPDFs presents noteworthy  similarity in shape among all cases. Also, the axis of the JPDF follows 
\begin{equation}
  \tilde{\varepsilon} D/U_{\infty}^3 = C_R Fr^{-1} {Ri}_g^{-1}, \label{kin_rel}
\end{equation}  {plotted as dashed lines in Fig.~\ref{jpdf_dissp_rig}.} 
 {Evidently, \eqref{kin_rel} provides a quantitative relationship -- a  power law -- between dissipation and $Ri_g$ across a range of stable and unstable $Ri_g$ and at different $Fr$. The following scaling analysis is useful to deduce \eqref{kin_rel}. }


 {In the limit of stable, strongly stratified flow, dissipation is predominantly from the vertical shear of horizontal velocity components, $\tilde{\varepsilon} \approx  \nu {S}_l^2$, where ${S}_l^2 = (\partial_z {u})^2 + (\partial_z {v})^2$ is the instantaneous, squared shear. The local stratification (${N}_l$) is close to the global constant $N$, leading to a gradient Richardson number of ${Ri}_g = N_l^2/{S}_l^2 \approx N^2/{S}_l^2$. The dissipation can then be expressed as}  {${\tilde{\varepsilon}} \approx \nu {S}_l^2 \approx \nu N^2 {Ri_g}^{-1}$.}   {In another limit of fully turbulent, nearly isotropic turbulence, $\tilde{\varepsilon} =  \nu (\partial_j  u_i)^2 \approx \frac{15}{2} \nu (\partial_z u)^2 = \frac{15}{4} \nu S_l^2$, invoking isotropic statistical relations. Although the proportionality constant is different in the two limits, the difference is small on a logarithmic scale and the power-law dependence remains the same.}
The inertially-normalized relation becomes
\begin{equation}
  \tilde{\varepsilon} D/U_{\infty}^3 \approx Fr_D^{-2} Re^{-1} {Ri}_g^{-1}, 
\end{equation}  {which simplifies to \eqref{kin_rel} with $Fr_D Re = Re_N =900$ being a constant and $Fr_D = Fr(h/D)$ being linearly proportional to the case-dependent $Fr$. }

Equation \eqref{kin_rel} serves as a reference for  {the expected instantaneous dissipation} in the context of stratified turbulence parameters. It can be seen in Fig.~\ref{jpdf_dissp_rig} that the JPDFs are very narrow around the dashed line given by \eqref{kin_rel} when $Ri_g$ is close to unity,  {corresponding to the stable limit. As $Ri_g$ decreases and the transition to instability occurs, the JPDFs widen  as there is scatter in dissipation.} The low-$Ri_g$  {spread of the JPDF of $\tilde{\varepsilon}$} is particularly evident at the larger $Fr =  0.3$ and $0.4$ values,  {where the same $Ri_g$ corresponds to $\tilde{\varepsilon}$ that differ by one to two orders of magnitude.} This is due to the intermittency of the local shear and stratification. As strong 3D turbulence is present,  {both larger local shear and lower local stratification (due to mixing of density) occur, and it becomes causally unclear whether lower $Ri_g$ leads to turbulence or turbulent overturning motions reduces $Ri_g$. But in all, \eqref{kin_rel} serves as a reasonable, quantitative prediction of the turbulent dissipation provided the local index $Ri_g$.} 

 {On the other hand, it is more challenging to predict dissipation using the local PV. The JPDFs of local dissipation and PV (not shown) display poor correlation -- given one, there is a wide scatter in the distribution of another. This is simply due to the fact that fully-developed turbulence contains intense vorticity and hence large PV (of both signs). While  large-scale PV in the base flow may imply (in)stability, small-scale PV in fully turbulent flow is  the result rather  than the cause of turbulence. 
}

\section{{Discussion and conclusion}} \label{theend}



Topographic features are ubiquitous on the seafloor  and are hot spots of turbulence generation. Both the  physical mechanisms that lead to turbulence and the accurate parametric dependence of dissipation on the overall governing parameters are crucial in the understanding and modeling of bottom ocean flows. To this end,   LES of the wake of an isolated 3D topography is employed for a cross-combination of four $Fr$ and five $Ro$, representing moderately strong to strong stratification and rotation rates that range from small MS to small SMS. The   LES is conducted at high resolution, sufficient to resolve flow instabilities and the energy-containing scales of wake turbulence. 
 {Our results indicate that  turbulence and dissipation in the near wake stem from two instabilities, KHI and CI, and the case-dependent intensity of dissipation depends on the development and manifestation of these instabilities.}
Volume-integrated dissipation in the near wake is quantified and rendered in the $(Fr,Ro)$ parameter space and, furthermore, the connection of this parametric dependence to the instability mechanisms is established. 

The primary instability in strongly stratified wakes is KHI between the dislocated layers  {of velocity--buoyancy structures}, which are oblique rather than horizontal.  
As stratification increases, the vertical length scale $U_{\infty}/N$ decreases and the distance between the layers becomes smaller (see Fig.~\ref{dissp_rig}). 
The velocity variation over such a distance, largely due to the out-of-phase vortex shedding at different heights, leads to intense shear collocated with the dislocation. The dimensional values of $\varepsilon$ as well as the spiky vertical profiles agree reasonably well with observational measurements. 
When stratification is relatively weaker, in cases Fr030 and Fr040, the hydraulic response becomes strong, which also breaks down to turbulence via KHI (see Fig.~\ref{dissp_rig}(g,h)). 

Rotation influences wake turbulence both indirectly and directly and the latter role depends on stratification. The indirect effect is through the modification of the coherent structures, which was comprehensively studied in \cite{liu2024effect}. When rotation is weak ($Ro>O(1)$), the vertical structures of wake vortices and dissipation are forward-slanted `surfboards' with $O(U_{\infty}/N)$ vertical thickness, as shown in Fig.~\ref{dissp_rig}(a,b), which is a configuration that  favors KHI. When rotation is strong ($Ro\le O(0.1)$), upright `columns' that are reminiscent of stratified Taylor columns are formed \citep{liu2024effect} and the vertical shear and KH turbulence are significantly reduced (see Fig.~\ref{dissp_rig}(e)). This indirect effect is also reflected in the volume-integrated dissipation rate for Fr007 and Fr015 cases, shown in Fig.~\ref{dissp_para}, that rotation appears to be solely stabilizing. The direct effect of rotation enters as stratification weakens, in cases Fr030 and Fr040, through CI. As the vertical length scale increases with increasing $Fr$, its constraint on CI is released and CI is able to destabilize the flow and change the dependence of dissipation on $Ro$ -- a dissipation peak emerges in the SMS range of $Ro=O(0.5)$. The existence of a most destabilizing rotation rate agrees with results from rotating horizontal shear layers \citep{yanase1993rotating,arobone2012evolution} {and circular flows} \citep{yim2016stability}, and can be theoretically expected as shown by the linear analysis in Appendix \hyperref[theory]{C}.

\begin{figure}[h!] 
  \centering 
  \captionsetup[subfloat]{farskip=0pt,captionskip=1pt}
  \subfloat{{\includegraphics[width=0.6\linewidth]{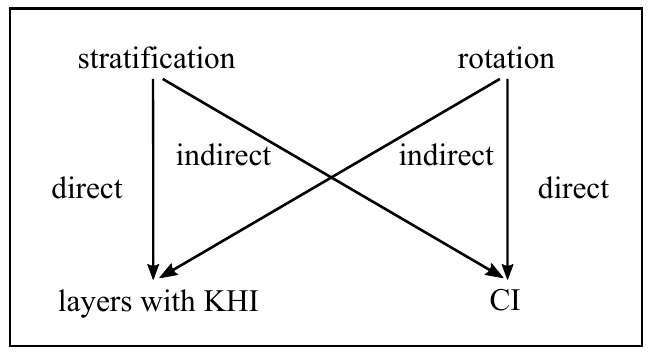}}} \hfill   \\ 
  \caption{Schematic showing the cross-influence of stratification and rotation on the KHI and CI in the present wakes. Arrows point to the direction of influence. 
  } \label{di-indi}
\end{figure}

To summarize, the instability mechanisms, KHI and CI, and the effects of stratification and rotation, are strongly  intertwined. The KHI results directly from the oblique dislocated layers due to strong stratification, but it is also indirectly influenced by rotation which alters the vertical shear. 
 {The CI has a non-monotonic dependence on rotation, while its growth is restricted by stratification.} 
It is the co-existence of KHI and CI, their subsequent nonlinear evolution and cross-dependence/influence that enriches the flow physics, while  {posing} challenges to simple parameterizations. {The idealized interaction between stratification and rotation in the context of wake turbulence  is shown as a schematic in Fig.~\ref{di-indi}}. 

Despite the variabilities as the instability diagnostics and integrated measures reveal, wakes at different $(Fr,Ro)$ share many characteristics in common. For example, the near wake box-integrated dissipation scales as   {$U^3_{\infty}D^2$}, and the scatter among different $(Fr,Ro)$ is within approximately a factor of three (see Fig.~\ref{dissp_para}). The JPDFs of $\tilde{\varepsilon}$ and $Ri_g$ shown in Fig.~\ref{jpdf_dissp_rig} share very similar shapes in all cases. The JPDFs are centered around $\tilde{\varepsilon} D/U_{\infty}^3 \sim Fr^{-1} Ri_g^{-1}$ but the distributions widen when $Ri_g$ is below its critical value 1/4, corresponding to the intermittency of the KHI. 


Toward the future  {research advancements} of wake turbulence and coherent eddies with LES studies, consideration of  more realistic geometry such as multiscale  topography or more complex seamount shapes is a promising direction.  Inclusion of realistic background flow that includes elements such as  non-linear stratification, non-uniform currents and strong tides would likely be very useful. 
{Topographic internal waves 
constitute an important ingredient of oceanic variability. Although the waves have induced dissipation that is weaker than wake turbulence (see Fig.~\ref{dissp_rig}) in the present problem, they propagate momentum and  energy into the ambient and, through their subsequent breakdown into turbulence,  present a reservoir  for remote topographic mixing of the ocean interior. The characteristics and parametric dependence of those waves  {in the context of 3D topography}, and their role 
in accomplishing remote mixing 
are subjects for future work. } 

\acknowledgments

This work is supported by ONR grant N00014-22-1-2024.


%
%
\datastatement {The reported data and corresponding {\rm MATLAB} scripts to read them are available in a public repository: \url{https://doi.org/10.5281/zenodo.15549518}.}








\clearpage

\appendix[A]  \label{sims} 
\appendixtitle{Detailed simulation setup} 

This appendix describes the details of the numerical setup, such as domain size, grid number, and resolution. Table \ref{table1b} provides the numerical setting.  

In the horizontal direction, the grid is dense near the hill (around $0.003\sim0.006D$ or $1.5\sim 3$ m) and is mildly stretched upstream and downstream. {The stretching ratio is reduced in the Fr040 cases, where the separation occurs further downstream, to resolve the gradients and control the numerical dispersion error due to grid stretching.} In the vertical direction, the resolution below $1.2h$ is kept at $15\sim 30$ grid points per $U_{\infty}/N$, in different cases, and there is also a mild stretching above. Since the stratification vertical length scale $U_{\infty}/N$ decreases as stratification increases, the number of vertical grid points  increases as $Fr^{-1}$.    {The boundary conditions on the seamount are  no-slip for velocity and no-flux for density. The flat bottom  is non-penetrating and shear-free, representing the scenario that the boundary layer thickness is small compared to the topography. The top boundary condition is Neumann, with a sponge layer that prevents reflection of internal waves back into the domain. The inflow condition is uniform velocity and linearly stratified, unperturbed density and the outflow is an Orlanski-type convective boundary that allows perturbations to freely propagate out of the domain without distortion and back reflection. }

\begin{table*}[t!]
  \begin{center}
  \begin{tabular}{c c c c c c c}
  \topline
    $Fr$  & $[N_x,N_y,N_z]$ & $[L_x, L_y, L_z]/D$ & $[\Delta x, \Delta y, \Delta z]/D$ in the NW & CPUs & CPU hours per case \\
    0.075 & $[1536,1152,432]$ & $[19, 7.6, 4.8]$ & $[0.0034,0.0066,0.0016]$ & 384 & 120 k \\
    0.15  & $[1536,1280,320]$ & $[19, 7.6, 4.2]$ & $[0.0034,0.0059,0.0024]$ & 256 & 75 k \\
    0.30  & $[1536,1280,216]$ & $[19, 7.6, 4.2]$ & $[0.0034,0.0059,0.0038]$ & 256 & 50 k \\
    0.40  & $[1920,1536,216]$ & $[16, 10, 4.2]$  & $[0.0034,0.0065,0.0038]$ & 384 & 60 k \\
  \botline
  \end{tabular}
  \caption{Simulation details. Overall, the largest near-wake  (NW) grid spacing is in the $y$-direction for  all cases, which is $\Delta y = 0.0066 D = 3.3 \,{\rm m}$ (with $D=500$ m). The vertical spacing is as small as  $\Delta z = 0.8$ m in the $Fr = 0.075$ series. The degree of freedom of the simulations ranges from $0.42$ to $0.76$ billion.   } \label{table1b}
  \end{center}
\end{table*}


A similar resolution was found to  provide good-quality LES in \cite{liu2024effect}. The approximate computational cost per case (production run) is listed in Table \ref{table1b} and the total cost is approximately {2 million CPU hours}, not counting the CPU time spent in pre-production scoping simulations.




\clearpage
\appendix[B]  
\appendixtitle{Effect of the Reynolds number} \label{re_sens}





At {sufficiently}  high Reynolds number, TKE dissipation is expected to be independent of viscosity. In equilibrium turbulence, the universal inertial scaling for dissipation $\langle \varepsilon \rangle = C_{\varepsilon} \mathcal{U}^3/\mathcal{L}$ is valid  {at large Reynolds number limit}.
 Here $\mathcal{U}$ is the fluctuating velocity scale and $\mathcal{L}$ is the integral length scale.  
 {At moderate Reynolds number, the `constant' slightly decays as $Re$ increases until the asymptote is reached} \citep{sreenivasan1984scaling,vassilicos2015dissipation}. 


 {In order to examine the sensitivity of our results to the Reynolds number,} three cases, Fr015Ro015, Fr015Ro075, Fr015Ro1p5, are selected for the $Re$-sensitivity  study. They are run at $Re=20\,000$ according to Table \ref{table1}. Simulations at two additional Reynolds numbers, $Re=10\,000, \, Re=30\,000$, are conducted for each of the three cases. 

\begin{figure*}[t!]
  \centering 
  \captionsetup[subfloat]{farskip=0pt,captionskip=1pt}
  \subfloat{{\includegraphics[width=0.32\linewidth]{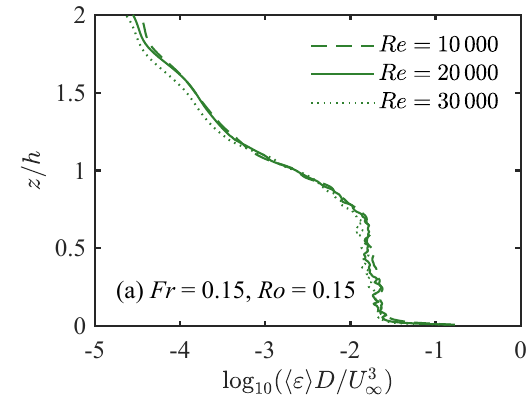}}} 
  \hspace{0cm} 
  \subfloat{{\includegraphics[width=0.32\linewidth]{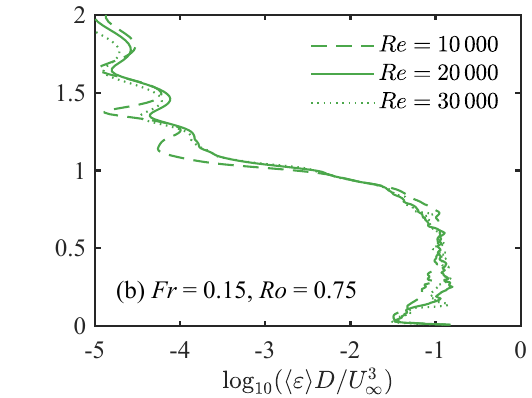}}}%
  \hspace{0cm} 
  \subfloat{{\includegraphics[width=0.32\linewidth]{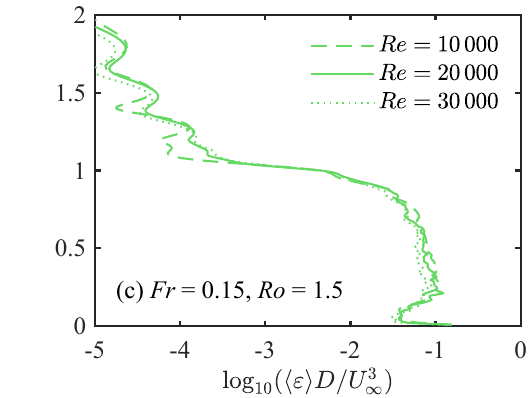}}}%
  \caption{TKE dissipation $\langle \varepsilon \rangle$ at various Reynolds numbers. 
  } \label{dissp_ReN} 
\end{figure*} 

\begin{figure*}[h!] 
  \centering 
  \captionsetup[subfloat]{farskip=0pt,captionskip=1pt}
  \subfloat{{\includegraphics[width=.5\linewidth]{./figs/tdissp_is_rig_clb.pdf}}} \hfill   \\ 
  \subfloat{{\includegraphics[width=0.45\linewidth]{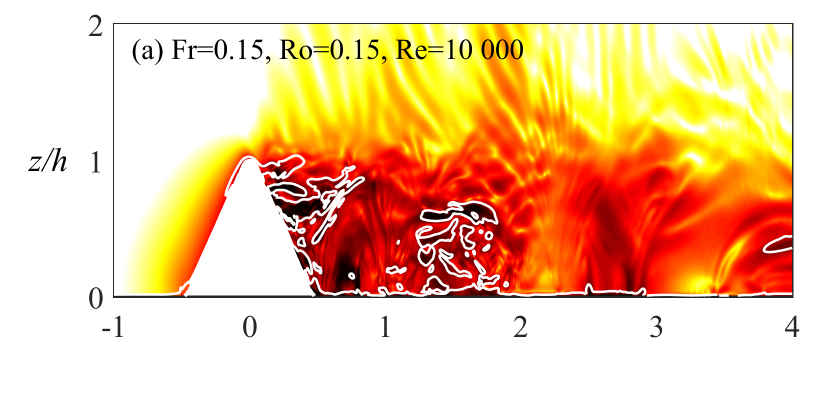}}}%
  \subfloat{{\includegraphics[width=0.45\linewidth]{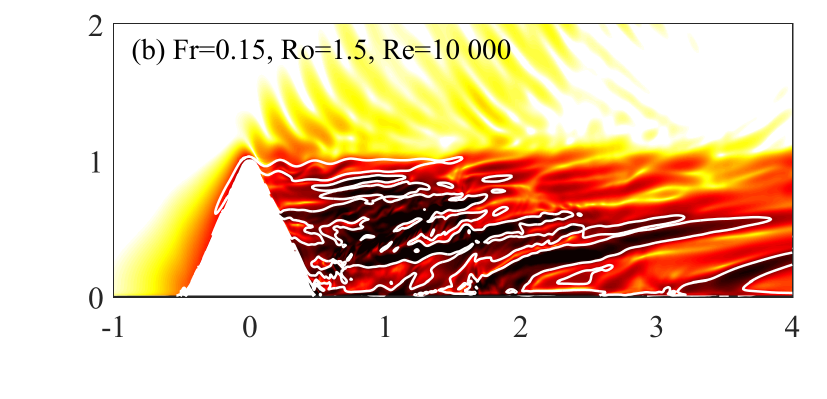}}}%
  \\
  \subfloat{{\includegraphics[width=0.45\linewidth]{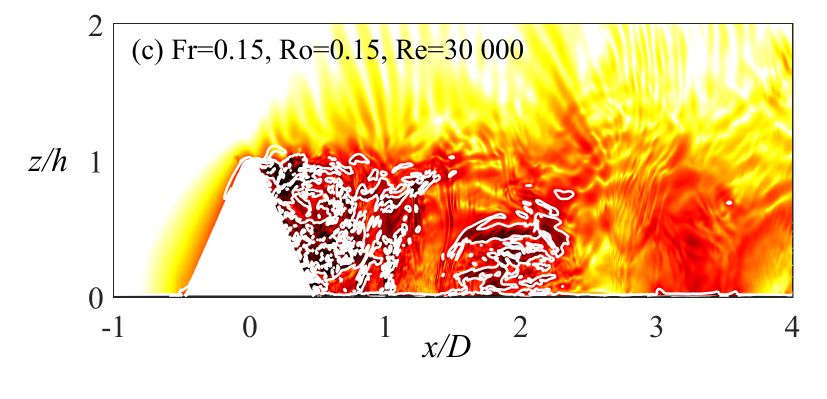}}}%
  \subfloat{{\includegraphics[width=0.45\linewidth]{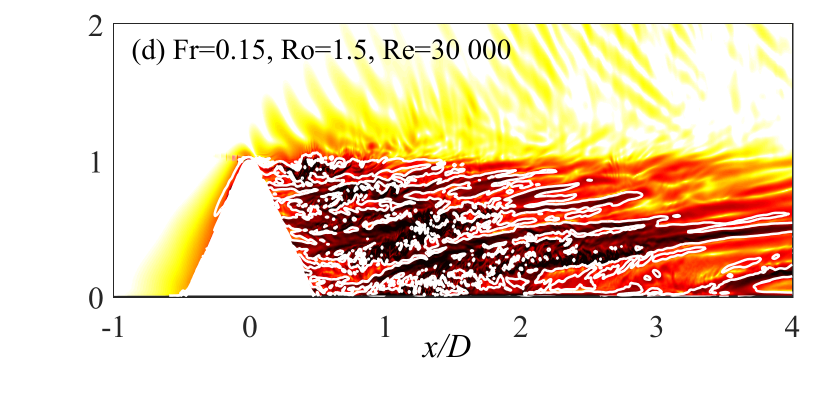}}}%
  \caption{Similar to Fig.~\ref{dissp_rig}. The Reynolds numbers are $Re=10\,000$ ($Re_N=450$) in (a,b) and $Re=30\,000$ ($Re_N=1350$) in (c,d). Panels (c,d) look very similar to Fig.~\ref{dissp_rig} (c-d), which are at $Re=20\,000$ ($Re_N=900$).  } \label{dissp_rig_re}
\end{figure*} 


In Fig.~\ref{dissp_ReN}, the time-averaged dissipation at $y=0,x/D=1$ is compared for three different Reynolds numbers. It can be seen that regardless of rotation, the dissipation rates at different Reynolds numbers agree reasonably well. In Fig.~\ref{dissp_ReN}(a,c), a slight decrease in dissipation as $Re$ increases is found, due to the moderate Reynolds number effect \citep{sreenivasan1984scaling,vassilicos2015dissipation}, but no additional instability mechanism or qualitative difference appeared. 

Figure \ref{dissp_rig_re} presents the contours of instantaneous dissipation in the center plane for cases Fr015Ro015 and Fr015Ro1p5, similar to Fig.~\ref{dissp_rig}(e,f) but at higher and lower Reynolds numbers. There are similarities of the large-scale structures between the Reynolds numbers, indicating that the large-scale coherent structures in stratified rotating wakes still remain robust despite higher Reynolds numbers. 
Meanwhile, there is a substantial addition of small scales to the flow {as $Re$ increases from 10,000 to 30,000} as a consequence of the breakdown of the instabilities. However, the change of the topology of the $\varepsilon$ field from $Re = 20,000$, which previously shown as Fig.~\ref{dissp_rig} (c-d),  to $Re = 30,000$ is small. This similarity and difference between the Reynolds numbers comply with the general picture of turbulence that the large scales are determined by the boundaries and forcing and the viscosity-dependent smallest scales become finer as the Reynolds number increases \citep{tennekes1972first}. 


Overall, the present Reynolds number is sufficiently high to trigger instabilities and the dissipation rate is relatively independent on Re. The differences among  cases with different $(Fr,Ro)$ are much more notable than those among different $Re$.


\appendix[C] 
\appendixtitle{{The inertial/Coriolis instability: a linear analysis}} \label{theory}


When a base flow with its own primary barotropic instability in the horizontal direction  {(for example, the inflection-point instability in jets or shear layers, or the K{\'a}rm{\'a}n shedding in bluff-body wakes)} is subject to  {unstable vertical rotation}, inertial/centrifugal/Coriolis (CI) modes emerge. These modes are baroclinic in the sense that they are usually characterized as horizontal vortex rollers \citep{kloosterziel2007inertial,arobone2012evolution,carnevale2013inertial}, which also lead to diapycnal overturns in a stratified fluid. {For completeness, in this Appendix, a linear analysis of CI is performed for two types of base flows: (1) parallel shear flow and (2) axisymmetric flow. 
The equations of horizontal vorticity perturbations are employed as in \cite{arobone2012evolution} for parallel shear flow and are generalized to curvilinear flow. It will be pointed out that conditions favorable for CI will lead to (initial) exponential growth of horizontal perturbation vorticities, and more specifically, at a growth rate of the square root of  respective stability criteria.  By examining the dependence of the initial growth rate on $Ro$, it is demonstrated that for both types of base flows the most destabilizing rotation rate falls roughly at $Ro=O(1)$.}

The starting point is the non-dimensional,  {linearized vorticity equation for inviscid perturbations:}  \begin{equation}
  \begin{aligned}
    \frac{\bar{D} \omega'_i}{\bar{D} t} = \omega'_j \langle S_{ij} \rangle + (\langle \omega_j \rangle + f_c \delta_{j3} ) S'_{ij} - u'_j \frac{\partial \langle \omega_i \rangle}{\partial x_j}  + \frac{1}{2}  \epsilon_{ij3} f_c \omega'_j + \epsilon_{ij3} Ri_b \frac{\partial \rho'}{\partial x_j} ,
  \end{aligned} \label{lin_vort}
\end{equation}  {obtained by taking the curl of the perturbed momentum equations linearized around a base flow. The base flow is denoted by brackets and the perturbations by primes.}  {The operator $\bar{D}/\bar{D}t = \partial_t + \langle u_i \rangle \partial_{x_i}$ represents mean convection and $S_{ij}$ is the rate-of-strain tensor.} The free indices $i=1,2,3$ denote the three spatial dimensions that are $(x,y,z)$ in Cartesian coordinates and $(r,\theta,z)$ in cylindrical coordinates. The corresponding velocity components are $(u,v,w)$. The base flow is $\langle u \rangle = U_{SL}(y)$ in the parallel flow case and is $\langle v \rangle = \langle v \rangle (r) = \langle u_{\theta}\rangle(r)$ in the axisymmetric flow case. The buoyancy Richardson number is $Ri_b = Fr_h^{-2}$, where $Fr_h$ is the horizontal Froude number, and the non-dimensional Coriolis parameter is $f_c=Ro^{-1}$.   



\subsection{Parallel shear flow} 

By solving linear stability equations, \cite{yanase1993rotating} and \cite{arobone2012evolution} showed that the eigenmodes of CI are typically streamwise-invariant ($\partial_x \approx 0$). With a {simplifying assumption $\partial_y w' \ll \partial_z v'$}, the $\omega'_x,\omega'_y$ components of \eqref{lin_vort} reduce to \begin{align}
  \frac{\partial \omega'_x}{\partial t} & = f_c \omega'_y - Ri_b \frac{\partial \rho'}{\partial y}, \\
  \frac{\partial \omega'_y}{\partial t} & = -(\langle \omega_z \rangle+f_c ) \omega'_x, 
\end{align} which form an autonomous system when the stratification is weak ($Ri_b\ll 1$): \begin{align}
  \frac{\partial }{\partial t} \begin{pmatrix}
    \omega'_x \\
    \omega'_y 
  \end{pmatrix} = & \begin{pmatrix}
    0 & f_c \\
    -(\langle \omega_z \rangle + f_c) & 0 
  \end{pmatrix} \begin{pmatrix}
    \omega'_x \\
    \omega'_y 
  \end{pmatrix}. \label{lin_sys} 
\end{align} 
Here, the background vorticity is $\langle \omega_z \rangle = - \partial_y \langle u \rangle = - 2 \langle S_{yx} \rangle$. The closed linear system \eqref{lin_sys} has two eigenvalues that satisfy \begin{equation}
  \lambda^2 = - f_c ( \langle \omega_z \rangle + f_c). \label{gr_car}
\end{equation} When \begin{equation}
   f_c ( \langle \omega_z \rangle + f_c) < 0, \label{abs_vor}
\end{equation} there is a pair of real eigenvalues of opposite signs, with the positive one $\lambda_+ = \sqrt{-f_c(f_c+\langle \omega_z \rangle )}$ corresponding to instability. Otherwise, a pair of purely imaginary, conjugate eigenvalues will imply inertial waves.

The condition \eqref{abs_vor} is equivalent to the absolute vorticity criterion for inertial instability \citep{holton1972introduction}. In particular, for a base flow with shear $\langle \omega_z \rangle$,  {the growth rate of the perturbations is the largest when $f_c = f_{c, {\rm max}} = -\langle \omega_z \rangle/2$, equivalently $Ro_{{\rm max}} = \langle \omega_z \rangle /f_{c, {\rm max}} = -2$ (anticyclonic)}. In other words, the most destabilizing  {rotation is   anticyclonic (with respect to the given shear), and is in the SMS range}.  {Cyclonic shear ($f_c$ with the same sign as $\langle \omega_z \rangle$) always result in stability according to} \eqref{gr_car}. When rotation is very strong compared to the shear ($Ro\ll O(1)$, $|f_c| \gg |\langle \omega_z \rangle|$),  $\lambda^2 = -f_c( \langle \omega_z \rangle + f_c) \approx -f_c^2<0$, the CI mode is stabilized  {regardless of the sign of $f_c$}, consistent with \cite{yanase1993rotating} and \cite{arobone2012evolution}. When rotation is very weak ($Ro\gg O(1)$, $f_c \approx 0$), 
 {the predicted growth rate \eqref{gr_car} vanishes since $f_c$ is small and $(\langle \omega_z \rangle + f_c)$ is finite. Thus, CI is stabilized at either strong or weak rotation. 
}  We also note that the optimal growth rate may not always be achieved, but nevertheless, it serves as a good predictor for the strength of CI.

 {The stability of the system \eqref{lin_sys} can be interpreted as the response of the shear layer to coordinate rotation. The system is the most resonant when the forcing frequency ($f_c$) is the closest to the intrinsic frequency of the system ($\langle \omega_z \rangle/2$, the angular velocity of the solid-body rotation part of fluid motions). The  {initial} amplification mechanism is  {linear}. 
 }

\subsection{Axisymmetric flow}

A similar linear theory can be established for axisymmetric flows, where $(u,v,w)$ and $(\omega_r, \omega_{\theta},\omega_z)$ denote the velocity and vorticity components in $r,\theta,z$ directions, respectively. The effect of stratification is similar to the parallel flow case and is not included at the outset. 

Similar to the quasi-streamwise-mode assumption in parallel flows, a quasi-axisymmetric-mode assumption ($\partial_{\theta} \approx 0$ for dependent variables; $\partial_{\theta}\bm{e}_r = \bm{e}_{\theta}$ and $\partial_{\theta} \bm{e}_{\theta} = -\bm{e}_{r}$ still apply) leads to \begin{equation}
  \begin{aligned}
    & \langle \bm{u} \rangle \cdot \nabla \bm{\omega}'  =   \frac{ \langle v \rangle}{r} \frac{\partial }{\partial \theta}  \left( \omega'_r \bm{e}_r + \omega'_{\theta} \bm{e}_{\theta} + \omega'_z \bm{e}_ z \right) \\ 
    & =  \left(\frac{\langle v \rangle}{r} \frac{\partial \omega'_r }{\partial \theta} - \frac{\langle v \rangle \omega'_{\theta}}{r} \right)\bm{e}_{r} + \left( \frac{\langle v \rangle}{r} \frac{\partial \omega'_{\theta}}{\partial \theta} + \frac{\langle v \rangle \omega'_r}{r} \right) \bm{e}_{\theta} 
    {+ \frac{\langle v \rangle}{r} \frac{\partial \omega'_{z}}{\partial \theta}  \bm{e}_{z}} \\
    & \approx - \frac{\langle v \rangle \omega'_{\theta}}{r} \bm{e}_{r} + \frac{\langle v \rangle \omega'_r}{r} \bm{e}_{\theta}. 
  \end{aligned}
\end{equation} Equation \eqref{lin_vort} is cast in cylindrical coordinates for $\omega'_r$ and $\omega'_{\theta}$ as \begin{align}
  \frac{\partial \omega'_r}{\partial t} - \frac{\langle v \rangle \omega'_{\theta}}{r}  = \omega'_{\theta} \langle S_{r \theta} \rangle + (\langle \omega_z \rangle + f_c) S'_{rz} + \frac{f_c}{2} \omega'_{\theta} \label{eqn_r} \\
  \frac{\partial \omega'_{\theta}}{\partial t} + \frac{\langle v \rangle \omega'_{r}}{r}  = \omega'_{r} \langle S_{ r \theta } \rangle + (\langle \omega_z \rangle + f_c) S'_{\theta z} - \frac{f_c}{2}\omega'_{r}.  \label{eqn_t}
\end{align} For axisymmetric base flow $\langle v \rangle(r)$, the mean rate-of-strain tensor and the mean vorticity are \begin{align}
  \langle S_{r \theta} \rangle  & = \frac{1}{2} \Biggl\langle r \frac{\partial }{\partial r} \left(
    \frac{v}{r}  \right) \Biggr\rangle = \frac{1}{2} \left( \frac{\partial \langle v \rangle}{\partial r} - \frac{\langle v \rangle }{r} \right); \;
    \langle \omega_z \rangle = \frac{\partial \langle v \rangle}{\partial r} + \frac{\langle v \rangle }{r}. 
\end{align} 
Furthermore, the quasi-axisymmetry also leads to $\omega'_r \approx - \partial_z v' = -2 S'_{\theta z}$ and the 
{simplifying assumption $\partial_r w' \ll \partial_z u'$} leads to $\omega'_{\theta} = 2 S'_{rz}$. 

The system \eqref{eqn_r}-\eqref{eqn_t} then takes an autonomous form \begin{equation}
  \frac{\partial}{\partial t} \begin{pmatrix}
    \omega'_r \\ \omega'_{\theta}
  \end{pmatrix} = \begin{pmatrix}
    0 & \langle \omega_z \rangle + f_c \\ 
    -\left( \frac{2 \langle v \rangle}{r} + f_c \right) & 0 
  \end{pmatrix} \begin{pmatrix}
    \omega'_r \\ \omega'_{\theta}
  \end{pmatrix}, 
\end{equation} with its eigenvalues satisfying \begin{equation}
  \lambda^2 = - \left( \frac{2 \langle v \rangle}{r} + f_c \right) (\langle \omega_z \rangle + f_c). \label{gr_cyl} 
\end{equation} The condition for instability is the eigenvalues being real, or \begin{equation}
  \chi = \left( \frac{2 \langle v \rangle}{r} + f_c \right) (\langle \omega_z \rangle + f_c) < 0, \label{chi}
\end{equation} where $\chi$ is exactly the generalized Rayleigh discriminant \citep{kloosterziel1991experimental,mutabazi1992gap}. Taking $\lambda_+(r)=\sqrt{-\chi(r)}$ as the estimated local growth rate for regions with $\chi<0$, the most destabilizing rotation rate can be searched for any axisymmetric eddy profile.
It is typical that an annular region satisfies $\chi<0$ and becomes unstable. Taking the derivative of \eqref{gr_cyl} with respect to $f_c$, the locally most destabilizing rotation rate  is $f_{c, {\rm max}}(r) = -(\langle v \rangle/r + \langle \omega_z \rangle /2)$. Although it should be evaluated with a global measure and it is flow-specific, the formal expression of $f_{c, {\rm max}}$ still suggests that  {an intermediate rotation rate at $Ro=O(1)$ will lead to the fastest three-dimensionalization}.   {It is also noted that, unlike in parallel flows, for which the instability criterion \eqref{abs_vor} can only be met with anticyclonic shear ($\langle \omega_z \rangle$ and $f_c$ have opposite signs), in axisymmetric flows both cyclonic and anticyclonic vortices can become unstable to CI. }

 {The implication of this Appendix is that the criteria \eqref{abs_vor} and \eqref{chi} are not only useful as stability discriminators, but they are also connected to how fast the perturbations can grow. The latter fact is utilized to predict the most destabilizing rotation rate for a given base flow, and has led to the understanding that intermediate rotation rates are the most unstable for CI, theoretically explaining the findings in section \ref{param}.}

\bibliographystyle{ametsocV6}
\bibliography{jjl}

\end{document}